\documentclass[10pt]{article}
\usepackage[OE]{express}

\usepackage[]{graphicx}
\usepackage{amssymb,amsmath,url,psfrag}
\usepackage{verbatim}
\usepackage{bm}
\usepackage{graphicx}
\usepackage{pstricks}
\usepackage{ulem} 
\usepackage{xcolor}

\def\added#1{#1}
\def\removed#1{}

\def\proofAdded#1{#1}
\def\proofRemoved#1{}

\newcommand{\Bolivarallee}{Boliva\hspace{-0.1mm}r\hspace{0.15mm}a\hspace{-0.1mm}llee}

\newcommand{\Takustrasse}{Taku\hspace{0.15mm}s\hspace{-0.1mm}tra{\ss}e}

\newcommand{\sketch}{
\def\svgwidth{0.3\linewidth}
\phantom{.}\hspace{0.7cm}
\begingroup%
  \makeatletter%
  \providecommand\color[2][]{%
    \errmessage{(Inkscape) Color is used for the text in Inkscape, but the package 'color.sty' is not loaded}%
    \renewcommand\color[2][]{}%
  }%
  \providecommand\transparent[1]{%
    \errmessage{(Inkscape) Transparency is used (non-zero) for the text in Inkscape, but the package 'transparent.sty' is not loaded}%
    \renewcommand\transparent[1]{}%
  }%
  \providecommand\rotatebox[2]{#2}%
  \ifx\svgwidth\undefined%
    \setlength{\unitlength}{467.01250211bp}%
    \ifx\svgscale\undefined%
      \relax%
    \else%
      \setlength{\unitlength}{\unitlength * \real{\svgscale}}%
    \fi%
  \else%
    \setlength{\unitlength}{\svgwidth}%
  \fi%
  \global\let\svgwidth\undefined%
  \global\let\svgscale\undefined%
  \makeatother%
  \begin{picture}(1,0.13717438)%
    \put(0,0){\includegraphics[width=\unitlength,page=1]{Sketch_spherical.pdf}}%
    \put(0.54784568,0.01731176){\color[rgb]{0,0,0}\makebox(0,0)[lb]{\smash{Spherical lens}}}%
  \end{picture}%
\endgroup%

\hspace{1.8cm}
\def\svgwidth{0.3\linewidth}
\begingroup%
  \makeatletter%
  \providecommand\color[2][]{%
    \errmessage{(Inkscape) Color is used for the text in Inkscape, but the package 'color.sty' is not loaded}%
    \renewcommand\color[2][]{}%
  }%
  \providecommand\transparent[1]{%
    \errmessage{(Inkscape) Transparency is used (non-zero) for the text in Inkscape, but the package 'transparent.sty' is not loaded}%
    \renewcommand\transparent[1]{}%
  }%
  \providecommand\rotatebox[2]{#2}%
  \ifx\svgwidth\undefined%
    \setlength{\unitlength}{467.01250211bp}%
    \ifx\svgscale\undefined%
      \relax%
    \else%
      \setlength{\unitlength}{\unitlength * \real{\svgscale}}%
    \fi%
  \else%
    \setlength{\unitlength}{\svgwidth}%
  \fi%
  \global\let\svgwidth\undefined%
  \global\let\svgscale\undefined%
  \makeatother%
  \begin{picture}(1,0.13717438)%
    \put(0,0){\includegraphics[width=\unitlength,page=1]{Sketch_mesa.pdf}}%
    \put(0.54566099,0.01668051){\color[rgb]{0,0,0}\makebox(0,0)[lb]{\smash{Mesa}}}%
  \end{picture}%
\endgroup%

}
\newcommand{\AlGaAs}{Al$_{0.9}$Ga$_{0.1}$As}
\begin{document}

\title{Numerical optimization of the extraction efficiency of a quantum-dot based single-photon emitter into a single-mode fiber}

\author{
Philipp-Immanuel~Schneider,\authormark{1}
Nicole~Srocka,\authormark{2}
Sven~Rodt,\authormark{2}
Lin~Zschiedrich,\authormark{1}
Stephan~Reitzenstein, \authormark{2}
and Sven~Burger\authormark{1,3}
}

\address{
\authormark{1}JCMwave GmbH,
\Bolivarallee~22, 
D\,--\,14\,050 Berlin,
Germany\\
\authormark{2}Technische Universit\"at Berlin,
Hardenbergstra{\ss}e~36,
D\,--\,10\,623 Berlin,
Germany\\
\authormark{3}Zuse Institute Berlin\,(ZIB),
\Takustrasse~7,
D\,--\,14\,195 Berlin,
Germany\\
}
\email{\authormark{1}philipp.schneider@jcmwave.com}

\begin{abstract}
We present a numerical method for the accurate and efficient simulation of strongly localized light sources, such as quantum dots, embedded in dielectric micro-optical structures. We apply the method in order to optimize the photon extraction efficiency of a single-photon emitter consisting of a quantum dot embedded into a multi-layer stack with further lateral structures. Furthermore, we present methods to study the robustness of the extraction efficiency with respect to fabrication errors and defects.
\end{abstract}


\bibliographystyle{osajnl}
\bibliography{bibl}

\begin{thebibliography}{10}
\newcommand{\enquote}[1]{``#1''}

\bibitem{Aharonovichetal.2016}
I.~Aharonovich, D.~Englund, and M.~Toth, \enquote{Solid-state single-photon
  emitters,} Nat. Photonics \textbf{10}, 631 (2016).

\bibitem{Purcell1946}
E.~M. Purcell, \enquote{Spontaneous emission probabilities at radio
  frequencies,} Proceedings of the American Physical Society \textbf{69}, 681
  (1946).

\bibitem{Barnes2002}
W.~Barnes, G.~Bj\"ork, J.~G\'erard, P.~Jonsson, J.~Wasey, P.~Worthing, and
  V.~Zwiller, \enquote{Solid-state single photon sources: light collection
  strategies,} Eur. Phys. J. D \textbf{18}, 197--210 (2002).

\bibitem{doi:10.1063/1.3245311}
A.~Muller, E.~B. Flagg, M.~Metcalfe, J.~Lawall, and G.~S. Solomon,
  \enquote{Coupling an epitaxial quantum dot to a fiber-based external-mirror
  microcavity,} Applied Physics Letters \textbf{95}, 173101 (2009).

\bibitem{doi:10.1063/1.3493187}
F.~Haupt, S.~S.~R. Oemrawsingh, S.~M. Thon, H.~Kim, D.~Kleckner, D.~Ding,
  D.~J.~S. III, P.~M. Petroff, and D.~Bouwmeester, \enquote{Fiber-connectorized
  micropillar cavities,} Applied Physics Letters \textbf{97}, 131113 (2010).

\bibitem{doi:10.1063/1.3617472}
M.~Davan\c{c}o, M.~T. Rakher, W.~Wegscheider, D.~Schuh, A.~Badolato, and
  K.~Srinivasan, \enquote{Efficient quantum dot single photon extraction into
  an optical fiber using a nanophotonic directional coupler,} Applied Physics
  Letters \textbf{99}, 121101 (2011).

\bibitem{Lee2015}
C.-M. Lee, H.-J. Lim, C.~Schneider, S.~Maier, S.~H{\"o}fling, M.~Kamp, and
  Y.-H. Lee, \enquote{Efficient single photon source based on m-fibre-coupled
  tunable microcavity,} Scientific Reports \textbf{5}, 14309 EP -- (2015).

\bibitem{doi:10.1063/1.4939264}
D.~Cadeddu, J.~Teissier, F.~R. Braakman, N.~Gregersen, P.~Stepanov, J.-M.
  Gérard, J.~Claudon, R.~J. Warburton, M.~Poggio, and M.~Munsch, \enquote{A
  fiber-coupled quantum-dot on a photonic tip,} Applied Physics Letters
  \textbf{108}, 011112 (2016).

\bibitem{Daveau:17}
R.~S. Daveau, K.~C. Balram, T.~Pregnolato, J.~Liu, E.~H. Lee, J.~D. Song,
  V.~Verma, R.~Mirin, S.~W. Nam, L.~Midolo, S.~Stobbe, K.~Srinivasan, and
  P.~Lodahl, \enquote{Efficient fiber-coupled single-photon source based on
  quantum dots in a photonic-crystal waveguide,} Optica \textbf{4}, 178--184
  (2017).

\bibitem{Somaschietal.2016}
N.~Somaschi, V.~Giesz, L.~De~Santis, J.~C. Loredo, M.~P. Almeida, G.~Hornecker,
  S.~L. Portalupi, T.~Grange, C.~Ant{\'o}n, J.~Demory, C.~G{\'o}mez, I.~Sagnes,
  N.~D. Lanzillotti-Kimura, A.~Lema{\'i}tre, A.~Auffeves, A.~G. White,
  L.~Lanco, and P.~Senellart, \enquote{Near-optimal single-photon sources in
  the solid state,} Nat. Photonics \textbf{10}, 340 (2016).

\bibitem{He:17}
Y.-M. He, J.~Liu, S.~Maier, M.~Emmerling, S.~Gerhardt, M.~Davan\c{c}o,
  K.~Srinivasan, C.~Schneider, and S.~H\"{o}fling, \enquote{Deterministic
  implementation of a bright, on-demand single-photon source with near-unity
  indistinguishability via quantum dot imaging,} Optica \textbf{4}, 802--808
  (2017).

\bibitem{Sapienzaetal.2015}
L.~Sapienza, M.~Davan{\c{c}}o, A.~Badolato, and K.~Srinivasan,
  \enquote{Nanoscale optical positioning of single quantum dots for bright and
  pure single-photon emission,} Nat. Communications \textbf{6}, 7833 (2015).

\bibitem{Badolato1158}
A.~Badolato, K.~Hennessy, M.~Atat{\"u}re, J.~Dreiser, E.~Hu, P.~M. Petroff, and
  A.~Imamo{\u g}lu, \enquote{Deterministic coupling of single quantum dots to
  single nanocavity modes,} Science \textbf{308}, 1158--1161 (2005).

\bibitem{Calic2017}
M.~Calic, C.~Jarlov, P.~Gallo, B.~Dwir, A.~Rudra, and E.~Kapon,
  \enquote{Deterministic radiative coupling of two semiconductor quantum dots
  to the optical mode of a photonic crystal nanocavity,} Scientific Reports
  \textbf{7}, 4100 (2017).

\bibitem{Gschreyetal.2015}
M.~Gschrey, A.~Thoma, P.~Schnauber, M.~Seifried, R.~Schmidt, B.~Wohlfeil,
  L.~Kr{\"u}ger, J.-H. Schulze, T.~Heindel, S.~Burger, F.~Schmidt,
  A.~Strittmatter, S.~Rodt, and S.~Reitzenstein, \enquote{Highly
  indistinguishable photons from deterministic quantum-dot microlenses
  utilizing three-dimensional in situ electron-beam lithography,} Nat.
  Communications \textbf{6}, 7662 (2015).

\bibitem{PhysRevB.81.125431}
Y.~Chen, T.~R. Nielsen, N.~Gregersen, P.~Lodahl, and J.~M\o{}rk,
  \enquote{Finite-element modeling of spontaneous emission of a quantum emitter
  at nanoscale proximity to plasmonic waveguides,} Phys. Rev. B \textbf{81},
  125431 (2010).

\bibitem{doi:10.1021/nl501648f}
G.~Bulgarini, M.~E. Reimer, M.~Bouwes~Bavinck, K.~D. J\"ons, D.~Dalacu, P.~J.
  Poole, E.~P. A.~M. Bakkers, and V.~Zwiller, \enquote{Nanowire waveguides
  launching single photons in a gaussian mode for ideal fiber coupling,} Nano
  Letters \textbf{14}, 4102--4106 (2014). PMID: 24926884.

\bibitem{doi:10.1021/acsphotonics.5b00559}
J.~Yang, J.-P. Hugonin, and P.~Lalanne, \enquote{Near-to-far field
  transformations for radiative and guided waves,} ACS Photonics \textbf{3},
  395--402 (2016).

\bibitem{Maes2013oe}
B.~Maes, J.~Petr\'{a}\v{c}ek, S.~Burger, P.~Kwiecien, J.~Luksch, and
  I.~Richter, \enquote{Simulations of high-{Q} optical nanocavities with a
  gradual {1D} bandgap,} Opt. Express \textbf{21}, 6794 (2013).

\bibitem{monk2003finite}
P.~Monk, \emph{Finite Element Methods for Maxwell's Equations}, Numerical
  Analysis and Scienti (Clarendon, 2003).

\bibitem{lavrinenko2014numerical}
A.~Lavrinenko, J.~L{\ae}gsgaard, N.~Gregersen, F.~Schmidt, and
  T.~S{\o}ndergaard, \emph{Numerical Methods in Photonics}, Optical Sciences
  and Applications of Light (CRC, 2014).

\bibitem{BERGOT2013189}
M.~Bergot and M.~Duruflé, \enquote{High-order optimal edge elements for
  pyramids, prisms and hexahedra,} Journal of Computational Physics
  \textbf{232}, 189 -- 213 (2013).

\bibitem{schlehahn2017stand}
A.~Schlehahn, S.~Fischbach, R.~Schmidt, A.~Kaganskiy, A.~Strittmatter, S.~Rodt,
  T.~Heindel, and S.~Reitzenstein, \enquote{A stand-alone fiber-coupled
  single-photon source,} arXiv preprint arXiv:1703.10536  (2017).

\bibitem{Kaganskiyetal.2017}
A.~Kaganskiy, S.~Fischbach, A.~Strittmatter, S.~Rodt, T.~Heindel, and
  S.~Reitzenstein, \enquote{Enhancing the photon-extraction efficiency of
  site-controlled quantum dots by deterministically fabricated microlenses,}
  Optics Communications \textbf{413}, 162 -- 166 (2018).

\bibitem{BurgerZschiedrichPomplunetal.2015}
S.~Burger, L.~Zschiedrich, J.~Pomplun, S.~Herrmann, and F.~Schmidt,
  \enquote{{Hp-finite element method for simulating light scattering from
  complex 3D structures},} Proc. SPIE \textbf{9424}, 94240Z (2015).

\bibitem{PhysRevLett.91.043902}
S.~M. Spillane, T.~J. Kippenberg, O.~J. Painter, and K.~J. Vahala,
  \enquote{Ideality in a fiber-taper-coupled microresonator system for
  application to cavity quantum electrodynamics,} Phys. Rev. Lett. \textbf{91},
  043902 (2003).

\bibitem{Novotny06principlesof}
L.~Novotny and B.~Hecht, \emph{{Principles of Nano-Optics}} (Cambridge
  University, 2006).

\bibitem{ZschiedrichGreinerBurgeretal.2013}
L.~Zschiedrich, H.~Greiner, S.~Burger, and F.~Schmidt, \enquote{{Numerical
  analysis of nanostructures for enhanced light extraction from OLEDs},} Proc.
  SPIE \textbf{8641}, 86410B (2013).

\bibitem{doi:10.1116/1.4914914}
M.~Gschrey, R.~Schmidt, J.-H. Schulze, A.~Strittmatter, S.~Rodt, and
  S.~Reitzenstein, \enquote{Resolution and alignment accuracy of
  low-temperature in situ electron beam lithography for nanophotonic device
  fabrication,} Journal of Vacuum Science \& Technology B, Nanotechnology and
  Microelectronics: Materials, Processing, Measurement, and Phenomena
  \textbf{33}, 021603 (2015).

\bibitem{SchneiderSantiagoetal.2017}
P.-I. Schneider, X.~Garcia~Santiago, C.~Rockstuhl, and S.~Burger,
  \enquote{{Global optimization of complex optical structures using Bayesian
  optimization based on Gaussian processes},} Proc.SPIE \textbf{10335}, 103350O
  (2017).

\bibitem{PhysRevLett.52.1798}
N.~Garcia and E.~Stoll, \enquote{Monte carlo calculation for
  electromagnetic-wave scattering from random rough surfaces,} Phys. Rev. Lett.
  \textbf{52}, 1798--1801 (1984).

\bibitem{nphys1870}
M.~Andersen, S.~Stobbe, A.~S{\o}rensen, and P.~Lodahl, \enquote{Strongly
  modified plasmon-matter inetraction with mesoscopic quantum emitters,} Nature
  Physics \textbf{7}, 215--218 (2010).

\bibitem{PhysRevLett.114.247401}
P.~Tighineanu, A.~S. S\o{}rensen, S.~Stobbe, and P.~Lodahl, \enquote{Unraveling
  the mesoscopic character of quantum dots in nanophotonics,} Phys. Rev. Lett.
  \textbf{114}, 247401 (2015).

\end{thebibliography}

\section{Introduction}

Single-photon emitters (SPE) are essential building blocks of future photonic and quantum optical devices. Solid-state SPEs, such as self-assembled quantum dots (QDs) and defect centers in solids, provide a scalable platform with outstanding optical properties in terms of the suppression of multiphoton emission and photon indistinguishability~\cite{Aharonovichetal.2016}. However, not only the properties of the SPEs but also the solid-state structures surrounding the QDs have an important influence on the system's performance. The surrounding structure can enhance or suppress the rate of spontaneous emission, a process known as Purcell effect~\cite{Purcell1946}. Moreover, the efficiencies with which emitted photons are extracted into a specific direction or coupled into \added{an optical fiber}\removed{a waveguide} depend strongly on the geometry of the surrounding structure~\cite{Barnes2002}.

\added{
So far, different strategies and structures were used to efficiently couple single-QD emission into an external waveguiding optical fiber in close-contact mode. Amongst others, fiber-based external-mirror microcavities~\cite{doi:10.1063/1.3245311}, DBR-based microcavities~\cite{doi:10.1063/1.3493187}, directional couplers~\cite{doi:10.1063/1.3617472}, photonic crystal micocavities~\cite{Lee2015}, nanowires directly attached to fibers~\cite{doi:10.1063/1.4939264}, and photonic crystal waveguides~\cite{Daveau:17} have demonstrated coupling efficiencies of up to 40\%~\cite{doi:10.1063/1.3493187} for QDs emitting below $980\,$nm.} With the aim to enhance the photon emission and extraction, deterministic nanoprocessing technologies have been developed and refined to integrate single QDs into micropillar cavities\cite{Somaschietal.2016,He:17}, circular dielectric gratings~\cite{Sapienzaetal.2015}, photonic crystal cavities~\cite{Badolato1158,Calic2017}, and microlenses~\cite{Gschreyetal.2015}.

Advanced nanoprocessing technologies allow for the realization of geometries with many degrees of freedom. However, from an experimental and technological perspective it is often not clear which geometries and geometric parameters lead to optimal results. In this case the numerical simulation of the micro-optical structures can give important insights~\cite{Gschreyetal.2015,PhysRevB.81.125431,doi:10.1021/nl501648f,doi:10.1021/acsphotonics.5b00559}.

In the following we present a method to simulate \removed{the properties of localized light sources} \added{the light field emitted by a QD} embedded into micro-optical structures \added{and its coupling efficiency to an external optical fiber}. 
\added{The coupling efficiency is very sensitive to the specific intensity and phase profile of the emitted light field when it enters the optical fiber. Hence, the numerical simulations require a high accuracy~\cite{Maes2013oe}. The presented numerical approach is based on the finite element method (FEM) in the frequency domain \cite{monk2003finite}. For nano-optical resonant devices, this method offers several important benefits compared to alternative methods, like  Finite Difference Time Domain (FDTD) or rigorous coupled-wave analysis (RCWA), with respect to accuracy and computation time. Firstly, the geometry is discretized with a non-uniform mesh allowing for an efficient and highly accurate modeling of the material interfaces without stair-casing effects as for FDTD and RCWA~\cite{lavrinenko2014numerical}. Secondly, the field is represented with local polynomial ansatz functions of various polynomial degree on the patches of the mesh. In this way and in contrast to standard FDTD, highly non-uniform field profiles can be captured by locally refined meshes, whereas regions of wave propagation can be very efficiently discretized using high-degree polynomial ansatz functions (p-refinement) \cite{BERGOT2013189}. Thirdly, since we are only interested in a QD emitting at a specific frequency, it is advantageous to solve Maxwell's equation directly in frequency domain. This is again in contrast to FDTD where the frequency response is reconstructed by a spectral analysis of a broadband excitation, which leads to long computation times especially for  fine meshes required for high accuracies. As a drawback, FEM has a higher memory footprint compared to FDTD. However, this was not the limiting factor for the studied SPE. In order to further improve the numerical convergence of the FEM approach, we exploit the radial symmetry of the system and use a subtraction method to cope with the singular behaviour of dipolar field emitted by the QD.}

We apply the \added{numerical} method in order to study \removed{the properties of} QDs embedded into micro-optical structures, which are fabricated on top of a Bragg reflector. 
The objective of the simulations is to identify geometries that efficiently couple the emitted light of the QD into \removed{a waveguide consisting of} an optical fiber above the structure \added{to further improve performance of stand-alone fiber-coupled single photon sources~\cite{schlehahn2017stand} in the future. }\removed{in order to facilitate user-friendly single-photon sources in the future.}
For clarity reasons, we focus on two embedding structure designs, microlenses and mesas, which can be produced with high accuracy utilizing 3D \emph{in-situ} electron-beam lithography~\cite{Gschreyetal.2015, Kaganskiyetal.2017}. 

The paper is organized as follows. After the introduction of the physical system of the embedded QD in Sec.~\ref{sec:System} we present the numerical approach in Sec.~\ref{sec:Method}. The approach is based on a finite element method (FEM)\cite{BurgerZschiedrichPomplunetal.2015} and
takes the highly singular nature of the QD light-field into account. We demonstrate that the computational times can be significantly reduced by exploiting the cylindrical symmetry of the considered geometry.
In Sec.~\ref{sec:Results} we apply the method in order to optimize the system parameters. By scanning the parameter space around the optima we assess the robustness of the geometry against the influence of fabrication errors.

\section{System description}
\label{sec:System}

In the following we consider a system consisting of a single QD emitting at a vacuum wavelength of $\lambda=1,300\,$nm in the telecom O-band. The QD is embedded into a spherical lens or a mesa structure made from gallium arsenide (GaAs; refractive index $n_{\rm struct} = 3.4$). An underlying Bragg reflector made from layers of GaAs ($n_{\rm Bragg1} = 3.4$) and aluminum gallium arsenide (\AlGaAs; $n_{\rm Bragg2}=3.0$) reflects the light emitted by the QD back into the upper hemisphere ($n_{\rm space} = 1.0$). The light is coupled into an optical fiber above the QD consisting of a homogeneous fiber core and a homogeneous fiber cladding ($n_{\rm core} = 1.5$, $n_{\rm clad} = 1.45$, ${\rm NA} \equiv \sqrt{n_{\rm core}^2-n_{\rm clad}^2} = 0.38$).

\begin{figure}[htp]
        \centering
        \def\svgwidth{0.45\linewidth}
        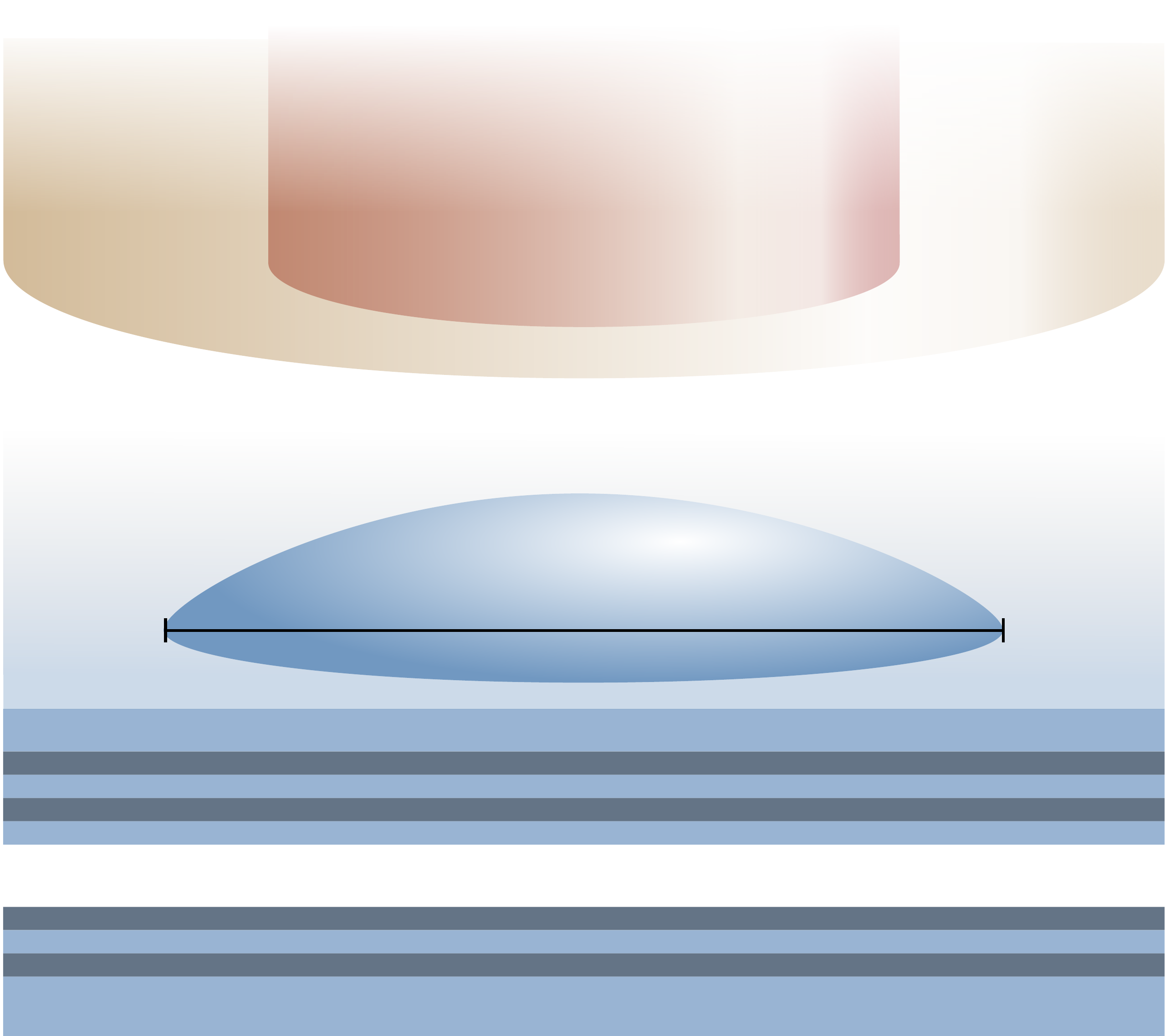
        \hfill
        \def\svgwidth{0.45\linewidth}
        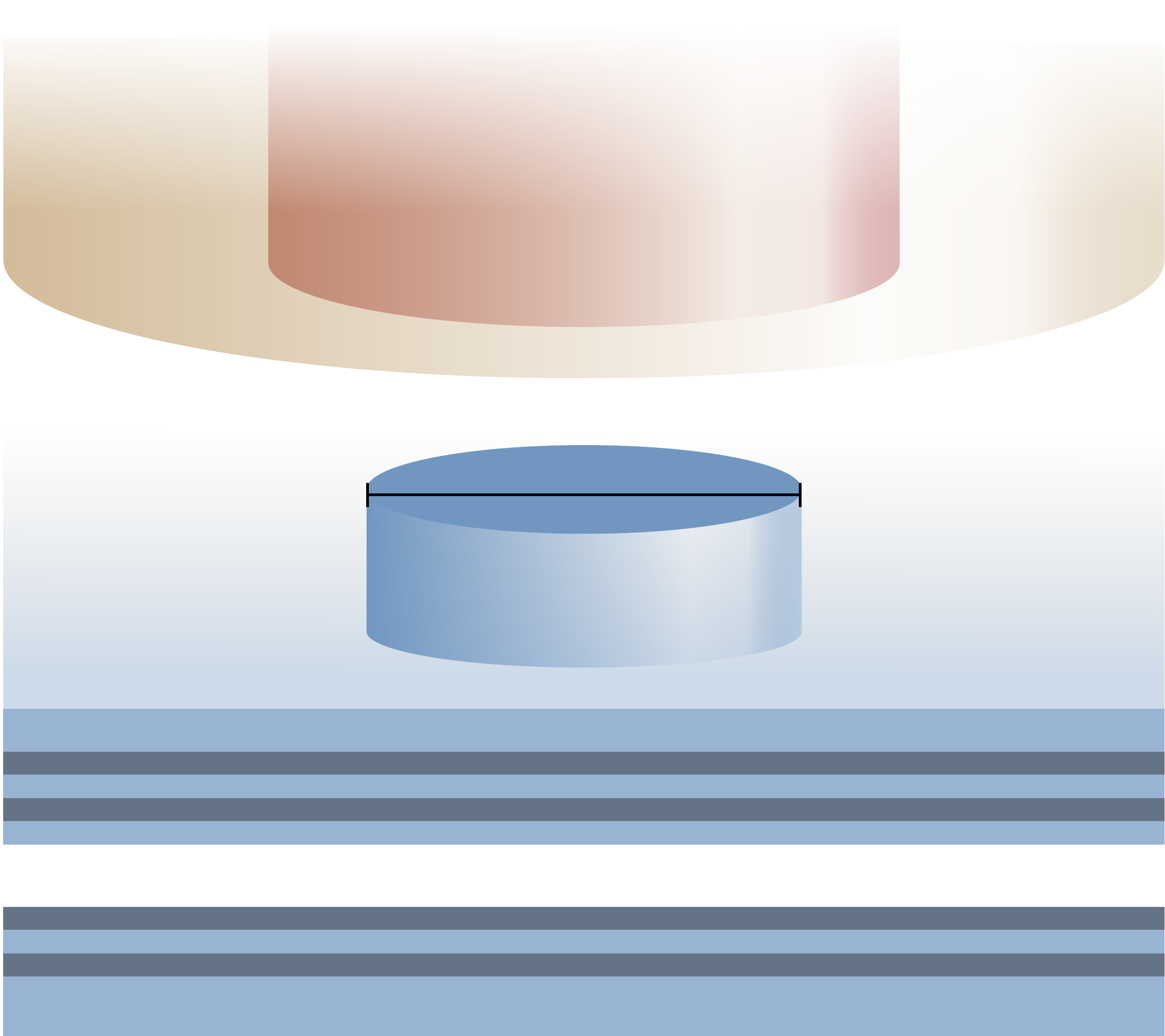
        \label{fig:setup_defect}\\

\caption{The considered system consists of a QD dipole source (red point) embedded into a diffractive structure (GaAs, blue), a Bragg reflector (alternating layers of GaAs (blue) and \AlGaAs~(gray)), and an optical fiber with homogeneous fiber core (orange) and fiber cladding (yellow). The Bragg reflector is grown on a substrate made of GaAs (blue) and has a GaAs top layer (blue, thickness \proofAdded{$=195\,$nm} \proofRemoved{$= 100\,$nm}).
The system is parametrized by 5 length scales, the diameter of the fiber core ($d_{\rm core}$), the width and height of the lens or mesa ($w_{\rm lens}$, $h_{\rm lens}$), the elevation of the dipole ($h_{\rm dip}$), and the distance between lens and fiber ($s_{\rm lf}$). \textbf{Left:} Spherical lens setup. \textbf{Right:} Mesa setup.}
\label{fig:setup}
\end{figure}

Figure~\ref{fig:setup} shows the two different types of diffractive structures -- spherical lens and mesa. The aim of these structures is to direct as much light as possible into the optical fiber. In Sec.~\ref{sec:Results} the system behavior will be studied as a function of  five geometry parameters (see Fig.~\ref{fig:setup}). These include the fiber-core diameter and the distance between fiber and spherical lens or mesa as well as the geometrical parameters of the embedding structure (dimensions of the spherical lens or mesa and the elevation of the QD). The geometry of the embedding structure can be controlled through the growth process and 3D \emph{in-situ} electron-beam lithography~\cite{Gschreyetal.2015}.

In the following, we consider fiber-core diameters between $1,000\,$nm and $2,500\,$nm. \added{Optical fibers with such small core diameters can be fabricated, e.~g., by flame heating and pulling standard fibers into a narrow thread~\cite{PhysRevLett.91.043902}.} \removed{such that the fiber supports only a single mode (or more exactly two degenerate modes with perpendicular polarization).} The mode field diameter (MFD) \added{of the single-mode fiber}, i.~e. the diameter where the field energy density has dropped to $1/e$ of the maximal field energy density, depends on the fiber-core diameter. For the considered parameter range the MFD varies between $2,400\,$nm and $3,400\,$nm. \added{The simulation results in Sec.~\ref{sec:Results} show that the relatively small fiber-core diameters and correspondingly small MFDs are favourable in order to gain a large overlap between the light field emitted by the QD and the propagation mode of the fiber.}

\section{Model and numerical method}
\label{sec:Method}

\subsection{Approximation of the quantum dot by a classical dipole}

The self-assembled QD under consideration consists of InGaAs and has a typical extension of $20\,$nm in horizontal direction and $5\,$nm in vertical direction. After excitation, it traps a single electron-hole pair, which can recombine through the spontaneous emission of a photon. To treat the interaction between the QD and the light field in the diffractive structure rigorously, one would need to invoke quantum electrodynamics (QED)~\cite{Novotny06principlesof}. However, the structure into which the photon is emitted has a low quality factor. That is, the photon energy diffuses quickly without having the time to re-interact with the QD. In this case one can model the emission properties of the QD by an oscillating point-like current density
\begin{equation}
\label{eq:J}
\mathbf{J}(\mathbf{r},t) = \added{
- {\rm Re}\left\{ i \omega \mathbf{p} \delta(\mathbf{r}- \mathbf{r_{\rm QD}}) e^{- i \omega t}\right\}}.
\end{equation}
Here, $\mathbf{r_{\rm QD}}$ is the position of the QD and $\omega$ its emission frequency. The QD has a larger lateral than vertical extension. Hence, the electronic state is excited horizontally and the dipole moment $\mathbf{p}$ lies in the horizontal plane. A more detailed explanation of the dipole approximation is given in Appendix A.

\subsection{Time-harmonic Maxwell's equations}

Since the considered source current density $\mathbf{J}(\mathbf{r},t)$ is time-harmonic (see Eq.~\eqref{eq:J}), the same holds for the electromagnetic fields $\mathbf{E}(\mathbf{r},t)$ and $\mathbf{H}(\mathbf{r},t)$. \removed{Therefore, the quantities can be expressed as ... [equations 2,3,4]}
\added{Therefore, we consider the} Maxwell's equations \added{for the electric field} in the frequency domain\added{~\cite{lavrinenko2014numerical}} \removed{can be cast into a second order form for the electric field}
\begin{equation}
\label{eq:time_harmonic}
\nabla\times\mu^{-1}\nabla\times \mathbf{E}(\mathbf{r},\omega) - \epsilon\omega^2  \mathbf{E}(\mathbf{r},\omega) = i\omega \mathbf{J}(\mathbf{r},\omega).
\end{equation}
\added{The} \removed{with the} permeability $\mu = \mu(\mathbf{r},\omega)$ and the permittivity $\epsilon = \epsilon(\mathbf{r},\omega)$ \added{depend on the material properties and the considered wavelength $\lambda = 2\pi/\omega$ of the light emitted by the QD}. \removed{The imaginary part $i\sigma/\omega$ of the permittivity tensor is proportional to the conductivity $\sigma(\mathbf{r},\omega)$ of the material.} %
For simplicity we will henceforth \removed{drop the hats and} implicitly assume that the fields and material tensors depend on $\omega$.
\added{The goal of the numerical method is to find the field distribution $\mathbf{E}(\mathbf{r})$ for different geometrical parameters of the system, and to determine in a second step the coupling efficiency of the field into the optical fiber.} 

\subsection{Subtraction method}

The electric field produced by the dipole emitter defined in Eq.~\eqref{eq:J} diverges for $\mathbf{r}\rightarrow\mathbf{r_{\rm QD}}$. Hence, a direct finite-element discretization of $\mathbf{E}$ suffers from a slow numerical convergence. To cure this, the electric field is expressed as a sum $\mathbf{E} = \mathbf{E}_s + \mathbf{E}_c$ of the singular dipole field $\mathbf{E}_s$ and a correction field $\mathbf{E}_c$~\cite{ZschiedrichGreinerBurgeretal.2013}. The singular field is chosen \added{to be the analytically known dipole-field solution of the Maxwell's equations} \removed{such that it is a solution of } 
\begin{equation}
\nabla\times\mu_d^{-1}\nabla\times \mathbf{E}_s(\mathbf{r}) - \epsilon_d \omega^2  \mathbf{E}_s(\mathbf{r}) = i\omega \mathbf{J}(\mathbf{r})
\end{equation} 
\added{with constant material tensors} \removed{where} $\mu_d = \mu(\mathbf{r_{\rm QD}})$ and $\epsilon_d = \epsilon(\mathbf{r_{\rm QD}})$ \removed{are constant}. %
\added{As shown in~\cite{ZschiedrichGreinerBurgeretal.2013}}%
\removed{Then}, the Maxwell's equation for the correction field reads\removed{[17]}
\begin{equation}
\label{eq:Ec}
\nabla\times\mu^{-1}\nabla\times \mathbf{E}_c(\mathbf{r}) - \epsilon \omega^2  \mathbf{E}_c(\mathbf{r}) = 
- \nabla\times(\mu^{-1} - \mu_d^{-1})\nabla\times \mathbf{E}_s(\mathbf{r}) + (\epsilon - \epsilon_d) \omega^2  \mathbf{E}_s(\mathbf{r}).
\end{equation}
\added{In the vicinity of the dipole emitter, the right-hand side of Eq.~\eqref{eq:Ec} vanishes since $\mu\rightarrow\mu_d$ and $\epsilon\rightarrow\epsilon_d$. Therefore, the correction field is smooth close to the dipole and can be efficiently computed using the finite-element method decribed in \cite{BurgerZschiedrichPomplunetal.2015}.}
\removed{This equation is solved with the finite-element method~[15].}

\subsection{Symmetry adaption}

For the considered setup, the material tensors are rotationally symmetric, i.~e. in cylindrical coordinates it holds $\mu(r,z,\phi)=\mu(r,z)$ and $\epsilon(r,z,\phi)=\epsilon(r,z)$ with $r=|\mathbf{r}|$. Therefore, we expand the electric field in a series of Fourier modes
\begin{equation}
\label{eq:series_Ec}
\mathbf{E}_c(\mathbf{r}) = \sum_{n=-\infty}^{\infty} \mathbf{E}_n (r,z) e^{i n \phi}.
\end{equation}
Likewise, the right hand side of Eq.~\eqref{eq:Ec},
\begin{equation}
\label{eq:series_f}
\mathbf{f}(\mathbf{r}) = - \nabla\times(\mu^{-1} - \mu_d^{-1})\nabla\times \mathbf{E}_s(\mathbf{r}) + (\epsilon - \epsilon_d) \omega^2  \mathbf{E}_s(\mathbf{r}),
\end{equation} can be expanded into a series
\begin{equation}
\mathbf{f}(\mathbf{r}) = \sum_{n=-\infty}^{\infty} \mathbf{f}_n (r,z) e^{i n \phi}\;\; \text{with}\;\; \mathbf{f}_n (r,z) = \frac{1}{2\pi}  \int_0^{2\pi} {\rm d}\phi\; \mathbf{f}(\mathbf{r})\, e^{- i n \phi}.
\end{equation}

Inserting Eqs.~\eqref{eq:series_Ec} and \eqref{eq:series_f} into Eq.~\eqref{eq:Ec} and integrating over $\frac{1}{2\pi}\int {\rm d}\phi\; e^{- i m \phi}$ yields independent 2-dimensional differential equations for every Fourier mode $m \in \mathbb{Z}$, i.~e.
\begin{equation}
\begin{split}
&\frac{1}{2\pi}\int_0^{2\pi}{\rm d}\phi e^{- i m\phi} \sum_{n=-\infty}^{\infty}\left(
\nabla\times\mu^{-1}\nabla\times \mathbf{E}_n(r,z) - \epsilon \omega^2  \mathbf{E}_n(r,z) - \mathbf{f}_n(r,z) \right) e^{i n \phi}\\
=& \sum_{n=-\infty}^{\infty}\left(
\widetilde{\nabla}\times\mu^{-1}\widetilde{\nabla}\times \mathbf{E}_n(r,z) - \epsilon \omega^2  \mathbf{E}_n(r,z) - \mathbf{f}_n(r,z) \right) \underbrace{\frac{1}{2\pi}\int_0^{2\pi}{\rm d}\phi\, e^{i (n-m)\phi}}_{= \delta_{n m}}\\
&= \widetilde{\nabla}\times\mu^{-1}\widetilde{\nabla}\times \mathbf{E}_m(r,z) - \epsilon \omega^2  \mathbf{E}_m(r,z) - \mathbf{f}_m(r,z) = 0,
\end{split}
\end{equation}
where $\widetilde{\nabla}$ arises from $\nabla$ by the replacement $\partial_\phi \rightarrow i n$. \added{Due to the reduced dimensionality of the equations the corresponding finite-element computations require much less computation time and reach a higher accuracy level than in the case Maxwell's equations are solved on a 3D mesh (see Sec.~\ref{ssec:convergence}).}

In numerical computations, only a finite number of Fourier modes is computed. The number of Fourier modes is chosen automatically by an adaptive algorithm such that the fields converge within some level of accuracy.
For the case that the dipole is positioned at the symmetry axis ($\mathbf{r}_{\rm QD} = (0,0,z)^T$), only the Fourier modes $m=-1,0,1$ contribute to the electric field expansion. However, we like to stress that even though the expansion relies on a cylindrical symmetric geometry, the source currents and light fields do not need to be symmetric. For example, also off-axis dipoles can be simulated by the method (see section~\ref{sec:Results}).

\subsection{Coupling efficiency to \removed{waveguide} \added{fiber} modes}

The geometry of the \removed{waveguide} \added{optical fiber} as described in section~\ref{sec:System} is invariant in $z$-direction. The eigenmodes $\mathbf{E}_n(\mathbf{r})$ for a given frequency $\omega$ are solutions of Eq.~\eqref{eq:time_harmonic} with no source current, i.~e.
\begin{equation}
\label{eq:mode}
\nabla\times\mu^{-1}\nabla\times \mathbf{E}_n(\mathbf{r}) - \epsilon\omega^2  \mathbf{E}_n(\mathbf{r}) = 0 \;\;\text{with}\;\;
\mathbf{E}_n(\mathbf{r}) = \mathbf{E}_n(r,\phi)e^{i k_n z}.
\end{equation}

For practical reasons one is interested in the fraction of the light field energy that is coupled into the fundamental mode of the \removed{waveguide} \added{optical fiber}. In order to define the coupling efficiency, we expand the electromagnetic field, which is scattered into the \removed{waveguide} \added{fiber}, into a series of \removed{waveguide} eigenmodes, i.~e., 
\begin{equation}
\mathbf{E}_{\rm scatt} = \sum_n \left<\mathbf{E}_n,\mathbf{E}_{\rm scatt}\right> \mathbf{E}_n.
\end{equation}

As scalar product we employ the following overlap integral:
\begin{equation}
\label{eq:scalar_prod}
\left<\mathbf{E}_1,\mathbf{E}_2\right> = \frac{1}{2 i \omega}\int {\rm d}\mathbf{n} \cdot (\mathbf{E}_1 \times\mu^{-1}\nabla\times \mathbf{E}_2) = \frac{1}{2}\int {\rm d}\mathbf{n} \cdot 
  (\mathbf{E}_1 \times \mathbf{H}_2),
\end{equation}
where the integration is performed over a cross section of the\removed{waveguide} \added{optical fiber}. 

In Appendix B we show that the \removed{waveguide} modes are orthonormal with respect to the scalar product of Eq.~\eqref{eq:scalar_prod}. Due to the orthonormality, the total power \added{$P = \frac{1}{2} {\rm Re} \left\{ \int \mathbf{n}\cdot (\mathbf{E}_{\rm scatt} \times \mathbf{H}^*_{\rm scatt})\right\}$} emitted into the \removed{waveguide} \added{optical fiber} is given as
\begin{equation}
\begin{split}
P &= \frac{1}{2} {\rm Re} \left\{ \int \mathbf{n}\cdot (\mathbf{E}_{\rm scatt} \times \mathbf{H}^*_{\rm scatt})\right\}
   = {\rm Re} \left\{ \left<\mathbf{E}_{\rm scatt},\mathbf{E}_{\rm scatt}^*\right>\right\} \\
  &= {\rm Re}\left\{ \sum_{n m} \left<\mathbf{E}_n,\mathbf{E}_{\rm scatt}\right>\left<\mathbf{E}_m,\mathbf{E}_{\rm scatt}^*\right> \left<\mathbf{E}_n,\mathbf{E}_m\right>\right\} = {\rm Re}\left\{\sum_{n} \left<\mathbf{E}_n,\mathbf{E}_{\rm scatt}\right>\left<\mathbf{E}_n,\mathbf{E}^*_{\rm scatt}\right>\right\}.\\
\end{split}
\end{equation}
In the case of guided modes in a loss-free non-active \removed{waveguide} \added{medium} \added{the eigenmodes $\mathbf{E}_n$ are real-valued and} the expression simplifies to
\begin{equation}
P= \sum_{n}  |\left<\mathbf{E}_n,\mathbf{E}_{\rm scatt}\right>|^2.
\end{equation}

The coupling efficiency towards a specific mode $n$ is defined as the powerflux $P_n = {\rm Re}\left\{\left<\mathbf{E}_n,\mathbf{E}_{\rm scatt}\right>\left<\mathbf{E}_n,\mathbf{E}^*_{\rm scatt}\right>\right\}$ to this mode divided by the total emitted power of the dipole $P_{\rm tot}$. In the following the coupling efficiency 
\begin{equation}
\label{eq:eta}
 \eta = \frac{P_1 + P_2}{P_{\rm tot}}
\end{equation}
to the two degenerate fundamental eigenmodes $\mathbf{E}_1$ and $\mathbf{E}_2$ of the \removed{waveguide} \added{optical fiber} is considered. \added{We would like to note, that the equations are not only valid for optical fibers, but for any waveguide structure.}

\subsection{Validation and convergence}
\label{ssec:convergence}
In order to validate the presented methods, we compare results of the symmetry-adapted simulations in effectively two dimensions (2D) with those of non-symmetry-adapted three-dimensional simulations (3D). To this end we consider a simplified setup consisting of a Bragg mirror with 10 double-layers ($n_{\rm Bragg1} = 3.4$, $n_{\rm Bragg2}=3.0$). The diffractive structure (spherical lens or mesa) is replaced by a flat GaAs layer with a thickness of \proofAdded{$395\,$nm} \proofRemoved{$200\,$nm} ($n_{\rm struct}=3.4$). The dipole is placed at \proofAdded{at a height of $195\,$nm within} \proofRemoved{the bottom of} this layer and the fiber with core diameter $1,500\,$nm ($n_{\rm core} = 1.5$, $n_{\rm clad} = 1.45$) has a distance of $100\,$nm to the flat GaAs layer ($n_{\rm space} = 1.5$). This setup has a coupling efficiency of $\eta=3.8\%$ to the two degenerate fundamental modes of the fiber. 

\begin{figure}
\includegraphics[width=\linewidth]{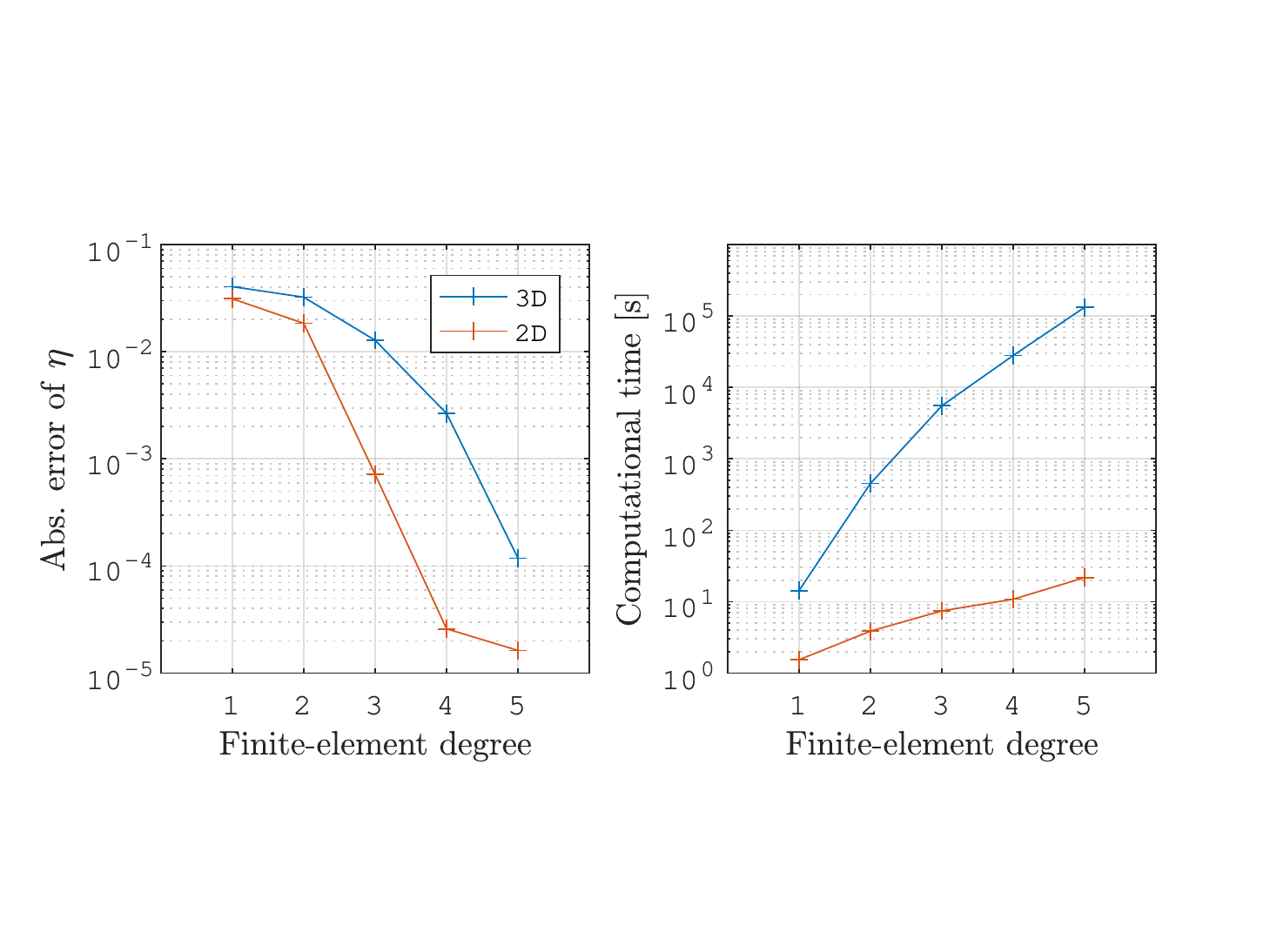} 
\caption{{\bf Left:} Convergence of the absolute error of the coupling efficiency as a function of the polynomial degree $p$ of the \added{finite-element ansatz functions (finite-element degree).} \removed{finite elements.} The 3D calculations (blue) converge significantly slower than the 2D calculations, which exploit the cylindrical symmetry. {\bf Right:} Computational time of the 2D and 3D calculations as a function of the polynomial degree of the finite elements. The converged 2D calculations are about 3 orders of magnitude faster than the 3D calculations.}
\label{fig:convergence}
\end{figure}

Figure~\ref{fig:convergence} shows the convergence behavior for an increase of the polynomial degree $p$ of the \added{finite-element ansatz functions (finite-element degree) \cite{BERGOT2013189}} \removed{finite elements} for the 2D and 3D computations. \added{Both computations converge to high numerical accuracies on the order of an absolute error of $10^{-4}$ showing that the methods provide reliable numerical results for the coupling efficiency.} In order to determine the coupling efficiency with an accuracy better than 1\% a polynomial degree of $p=3$ (2D case) or $p=4$ (3D case) is sufficient. \added{For this accuracy level the computation times of the 2D and 3D case differ drastically. While the 3D calculations require about 1 hour with 8 CPU threads, the 2D computation requires less than 10 seconds.} We attribute the slower convergence of the 3D calculations to the fact that the meshing of the structure in the 3D calculations leads to an additional discretisation error along the angular coordinate $\phi$. 

\removed{By exploiting the cylindrical symmetry, the computational times can be reduced by roughly three orders of magnitude for the described setup and for typical accuracy requirements.} In the next section, \added{the speed-up due to the exploitation of the cylindrical symmetry} \removed{this speed-up} will enable the exploration of the behavior of the coupling efficiency for a large number of different parameter values of the geometry. 

\section{Numerical results}
\label{sec:Results}

\subsection{Numerical optimization of the geometry}

In the following, we seek to optimize the geometry of the micro-optical structures such that the light emitted by the QD is efficiently coupled into the two fundamental modes of the fiber. The free parameters under consideration are the width $w_{\rm lens}$ and height $h_{\rm lens}$ of the diffractive structures (spherical lens or mesa), the elevation of the dipole $h_{\rm dip}$ within these structures, the distance between lens and fiber $s_{\rm lf}$, and the diameter of the fiber core $d_{\rm core}$ (see figure~\ref{fig:setup}).

\added{The numerical computations of the spherical lens and mesa geometries where performed on a computational domain with a radius of $6,000\,$nm and transparent boundary conditions. The systems were discretized with about $3,700$ triangular mesh elements and required computation times of about 40 seconds.}

\begin{figure}
\includegraphics[width=0.45\linewidth]{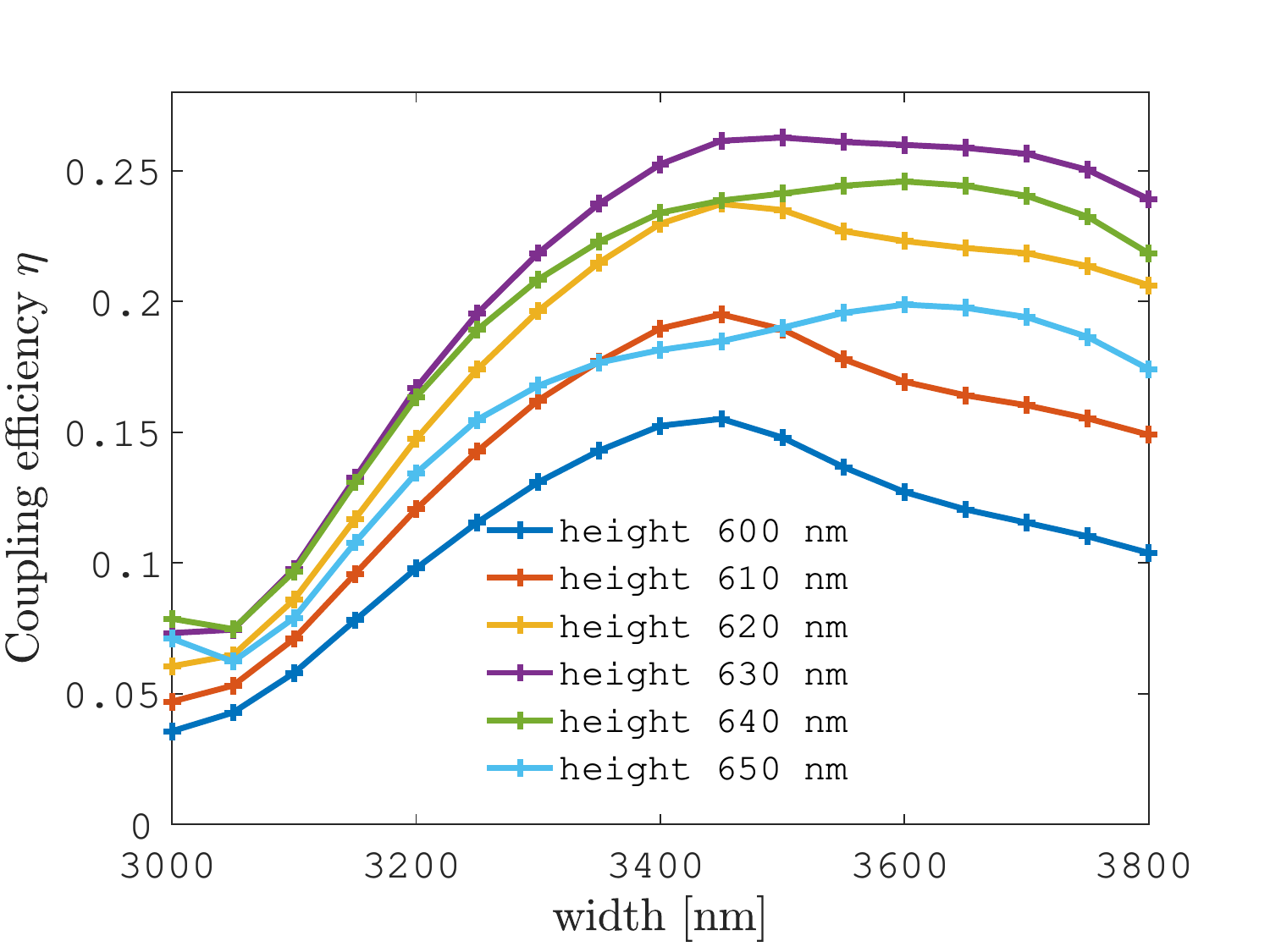} 
\includegraphics[width=0.45\linewidth]{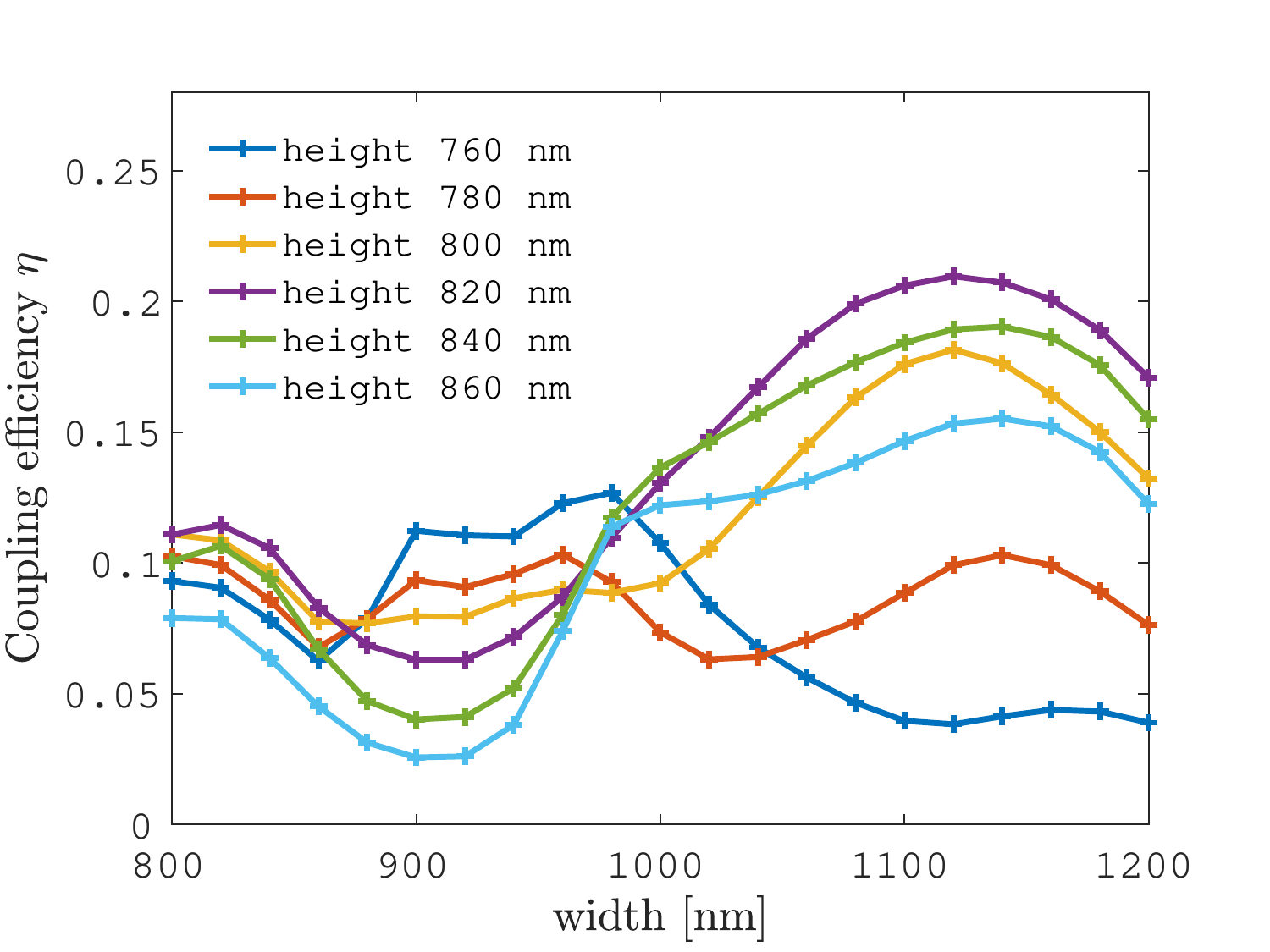} \vspace*{-4.8cm}
\\ 

\def\svgwidth{0.3\linewidth}
\phantom{.}\hspace{0.7cm}
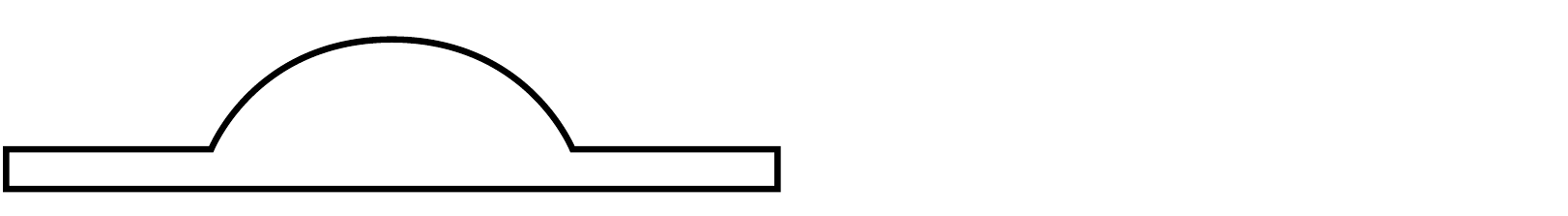
\hspace{1.8cm}
\def\svgwidth{0.3\linewidth}
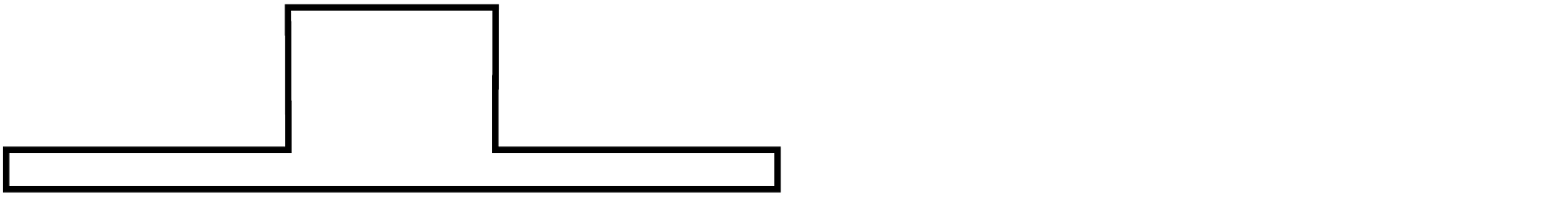

\vspace*{4.2cm}
\caption{Maximum coupling efficiency $\eta$ into a single-mode fiber as defined in Eq.~\eqref{eq:eta} for different widths and heights of a spherical lens (left) and a mesa (right) at a wavelength of $1,300\,$nm. The maximum is taken with respect to a scan of the dipole elevation within the range of 0 to $50\,$nm in steps of 10~nm. The values of the fiber core diameter and the lens-fiber distance are $d_{\rm core} = 1,500\,$nm and $s_{\rm lf} = 100\,$nm, respectively. \added{The best coupling efficiencies obtained in this optimization step are $\eta = 26.2\%$ for the spherical lens setup and $\eta = 21.0\%$ for the mesa setup}}
\label{fig:lower_optimization}
\end{figure}

The short computation times of the 2D calculations allow for a rather fine sampling of the parameter space. However, a sampling on a 5-dimensional grid would still require too many computations. Therefore, we first sample the parameters of the lower part of the setup ($w_{\rm lens}$, $h_{\rm lens}$, and $h_{\rm dip}$) for fixed parameters $d_{\rm core} = 1,500\,$nm and $s_{\rm lf} = 100\,$nm. The results of the parameter scans are shown in Fig.~\ref{fig:lower_optimization}. \added{Obviously, the coupling efficiency depends very sensitively on the width and height of the spherical lens and mesa. Especially the complex behavior for a variation of the width of the mesa is an indication that several resonant states within the structure can be exited, each having a different overlap with the fundamental propagation modes of the optical fiber.} In this step the best coupling efficiencies obtained are $\eta = 26.2\%$ for the spherical lens setup and $\eta = 21.0\%$ for the mesa setup. 

\begin{figure}
\includegraphics[width=0.45\linewidth]{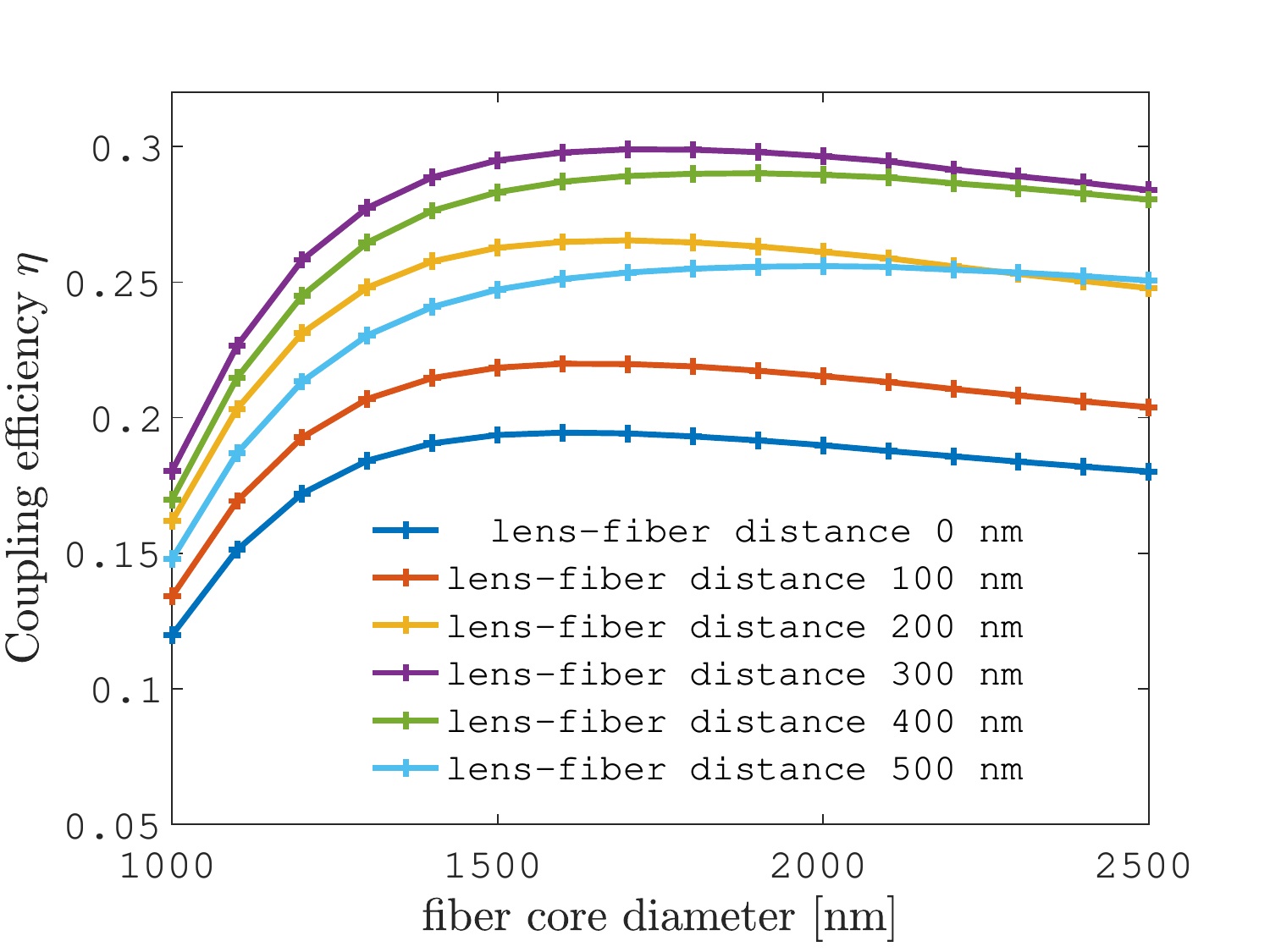} 
\includegraphics[width=0.45\linewidth]{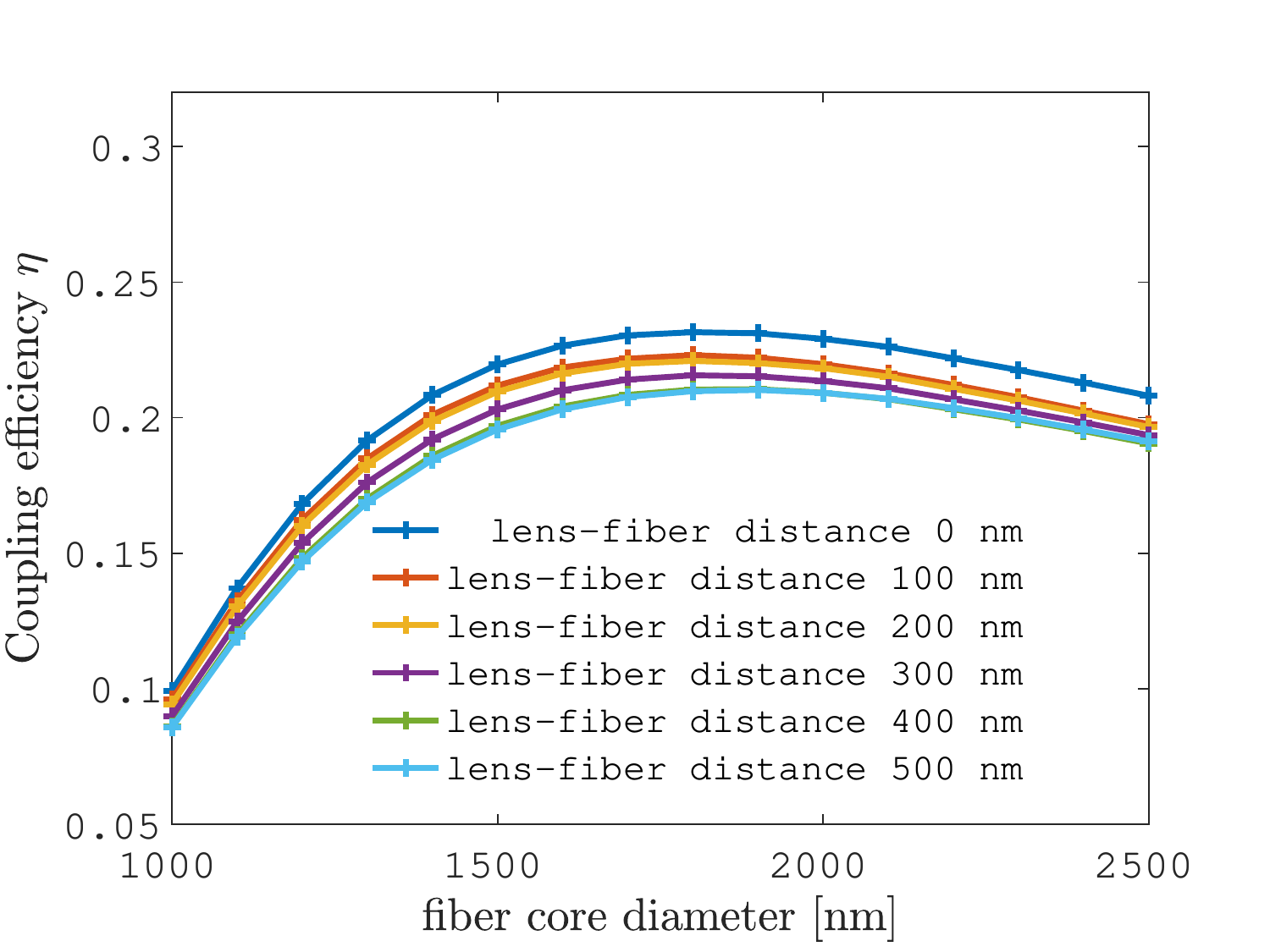} 
\vspace*{-4.8cm}
\\

\def\svgwidth{0.3\linewidth}
\phantom{.}\hspace{0.7cm}
\input{Sketch_spherical.pdf_tex}
\hspace{1.8cm}
\def\svgwidth{0.3\linewidth}
\input{Sketch_mesa.pdf_tex}

\vspace*{4.2cm}
\caption{Coupling efficiency $\eta$ into a single-mode fiber as defined in Eq.~\eqref{eq:eta} for different fiber core diameters and lens-fiber distances to the spherical lens (left) and mesa (right) at a wavelength of $1,300\,$nm. \added{The best coupling efficiencies obtained are $\eta = 29.9\%$ for the spherical lens setup and $\eta = 23.2\%$ for the mesa setup.} The other system parameters are set to the values of the first optimization step \added{(see Fig.~\ref{fig:lower_optimization})} \removed{(see Tab.~\ref{tab:optimal_parameters})}.
\added{The optimized parameters are summarized in Tab.~\ref{tab:optimal_parameters}}
}
\label{fig:upper_optimization}
\end{figure}

In the second step, the diameter of the fiber core $d_{\rm core}$ and the distance between lens and fiber $s_{\rm lf}$ are sampled while setting the values of $w_{\rm lens}$, $h_{\rm lens}$, and $h_{\rm dip}$ to the values  optimized in the first step. The results of the parameter scans are shown in Fig.~\ref{fig:upper_optimization}. \added{ The spherical lens setup is more sensitive on the lens-fiber distance than the mesa setup, indicating that the field radiated upwards has a stronger divergent behavior. For large fiber-core diameters the coupling efficiency of both structures slowly decreases showing that narrow tapered fibers are crucial in order to reach high coupling efficiencies. }After \added{the second optimization }\removed{this} step the best coupling efficiencies obtained are $\eta = 29.9\%$ for the spherical lens setup and $\eta = 23.2\%$ for the mesa setup. The optimized parameters are summarized in Tab.~\ref{tab:optimal_parameters}.

\begin{figure}
\includegraphics[width=0.5\linewidth]{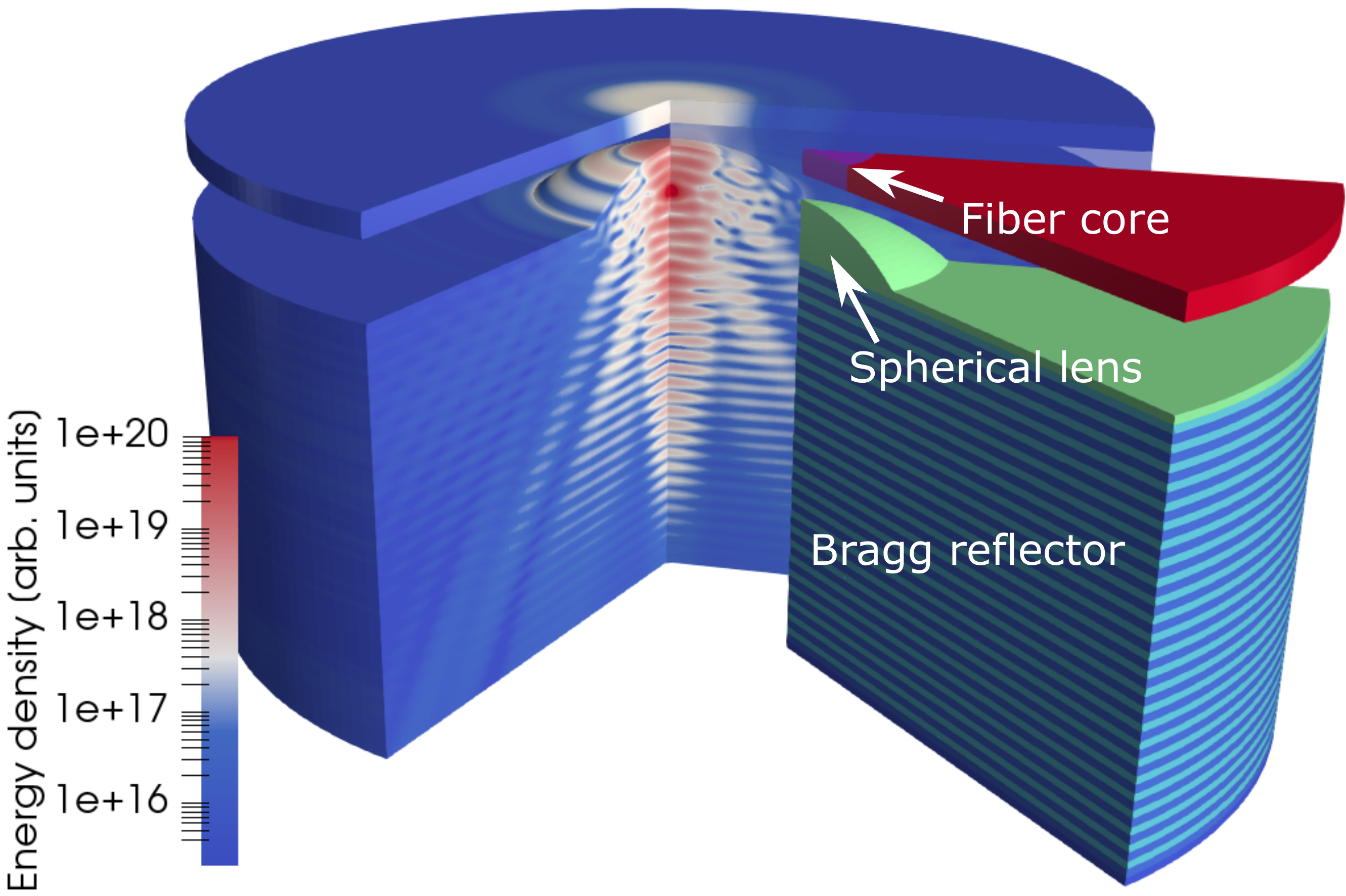} \hfill
\includegraphics[width=0.5\linewidth]{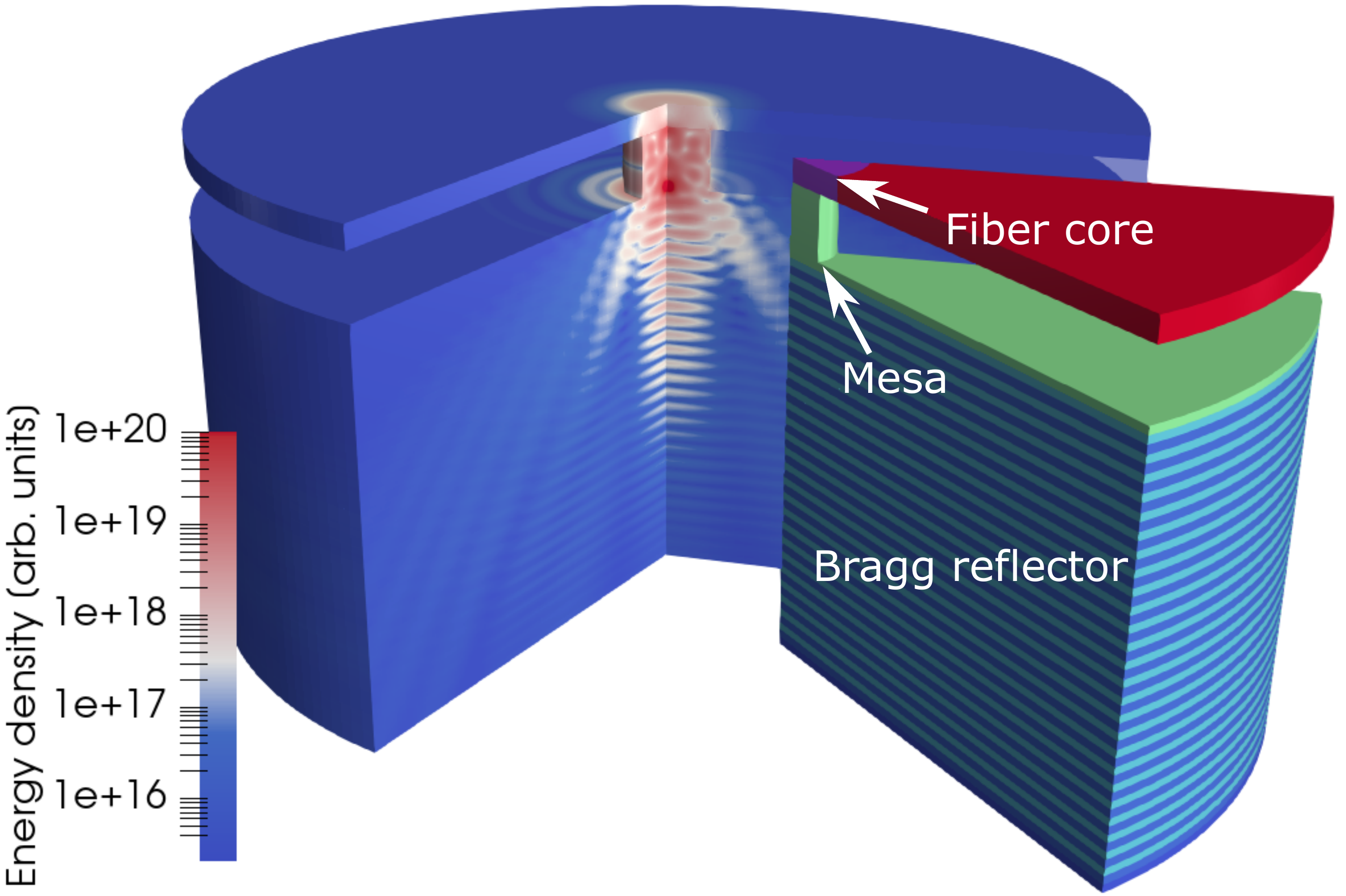} 

\caption{Visualization of the energy density \added{on a logarithmic scale} of the light field emitted by the QD embedded in the optimized spherical lens (left) and mesa (right). A cut through the optimized geometry is shown in front of the energy-density plots with the fiber core in purple, the fiber cladding in red, the spherical lens or mesa structure in green and the layers of the Bragg reflector in light and dark blue. \added{The optimal mesa has a smaller coupling efficiency although less radiation is propagating into the Bragg reflector. This indicates that the specific intensity and phase profile of the light field entering the fiber is of importance.}}
\label{fig:fields}
\end{figure}

\begin{table}[ht]
\caption{Optimized parameters and corresponding coupling efficiency $\eta$ for spherical lens and mesa obtained by a two-step optimization.}
\label{tab:optimal_parameters}
\begin{center}
\begin{tabular}{|p{7em}| r r r r r | p{6em}| }
\hline 
Diffractive structure & $w_{\rm lens}$ & $h_{\rm lens}$ & $h_{\rm dip}$ & $s_{\rm lf}$ & $d_{\rm core}$ & Coupling efficiency $\eta$ \\
\hline 
\hline 
Spherical lens & $3,500$\,nm & $630$\,nm & $30$\,nm & $300$\,nm & $1,700$\,nm & $29.9\%$ \\
Mesa lens & $1,120$\,nm & $820$\,nm & $30$\,nm & $0$\,nm & $1,800$\,nm & $23.2\%$ \\
\hline 
\end{tabular}
\end{center}
\end{table}

\added{In Fig.~\ref{fig:fields} the energy density distribution is visualized for the optimal geometries of the spherical lens and the mesa setups. For the optimal mesa setup less radiation is propagating into the Bragg reflector. Nevertheless, the spherical lens leads to a considerably larger coupling efficiency of about 30\% as compared to 23\% for the setup with the mesa. This indicates that the specific intensity and phase profile of the light field entering the fiber is of importance. }

\added{Although the spherical lens setup is performing better, the structure }
\removed{Clearly, within the considered part of the parameter space the setup with a spherical lens leads to a considerably larger coupling efficiency of about 30\% as compared to 23\% for the setup with the mesa. However, a spherical lens} is much harder to produce than a mesa. Moreover, small errors in the fabrication process can have a large influence on the geometry of the spherical lens. Therefore, in the next section we study the sensitivity of the coupling efficiency with respect to fabrication errors.

\subsection{Sensitivity analysis}

The fabrication of the diffractive structures is subject to various errors.	While the heights of the epitactically grown structures can be produced with a high accuracy of a few nanometers, the lateral accuracy is generally lower. For example, the lateral position of the dipole has an uncertainty of about $34\,$nm~\cite{doi:10.1116/1.4914914} and the widths of the spherical lens and mesa can be controlled with an accuracy of about $100\,$nm. Moreover, also the lens-fiber distance has a relatively large  uncertainty of about $50\,$nm.

\begin{figure}
\includegraphics[width=0.45\linewidth]{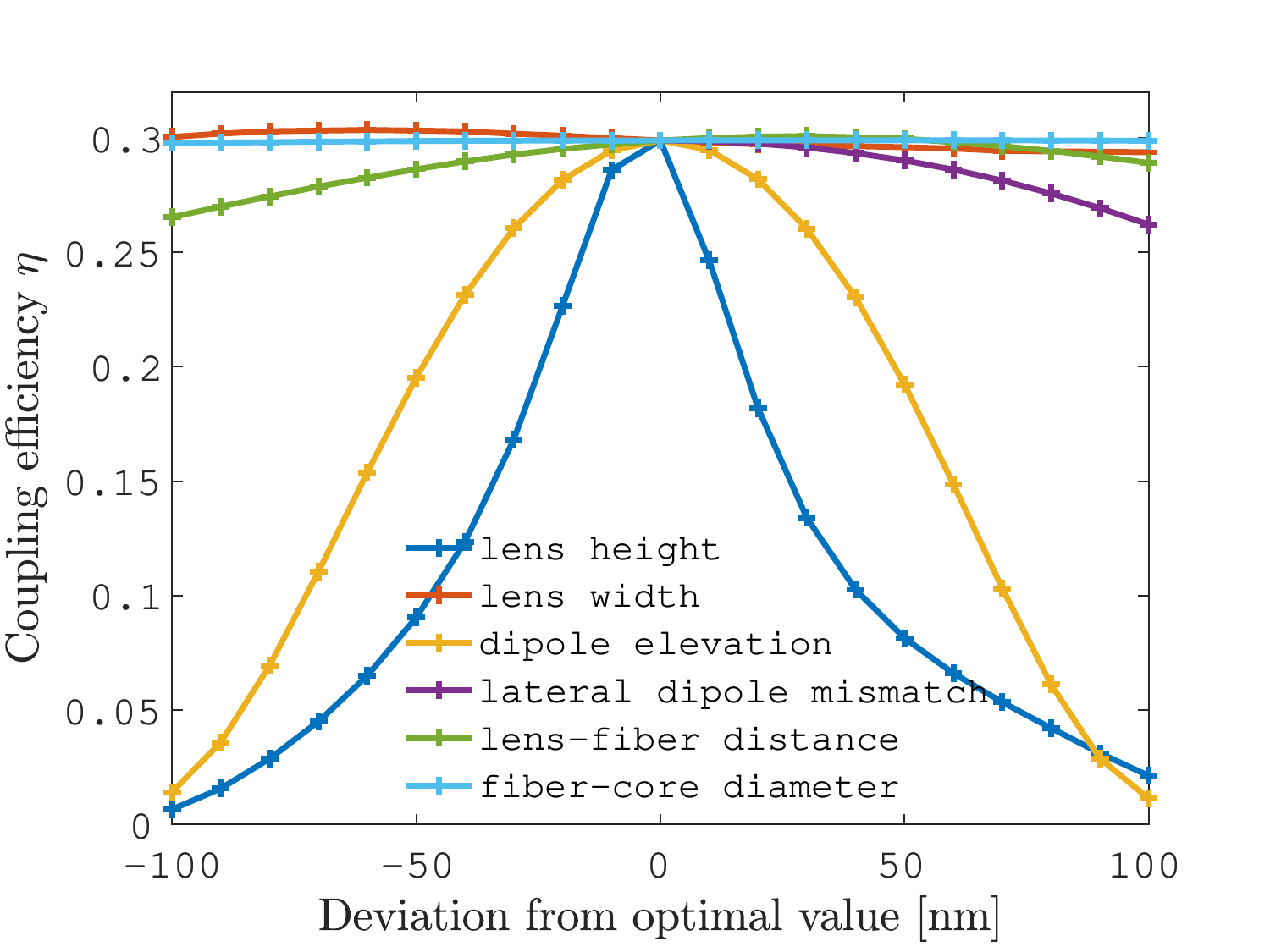} 
\includegraphics[width=0.45\linewidth]{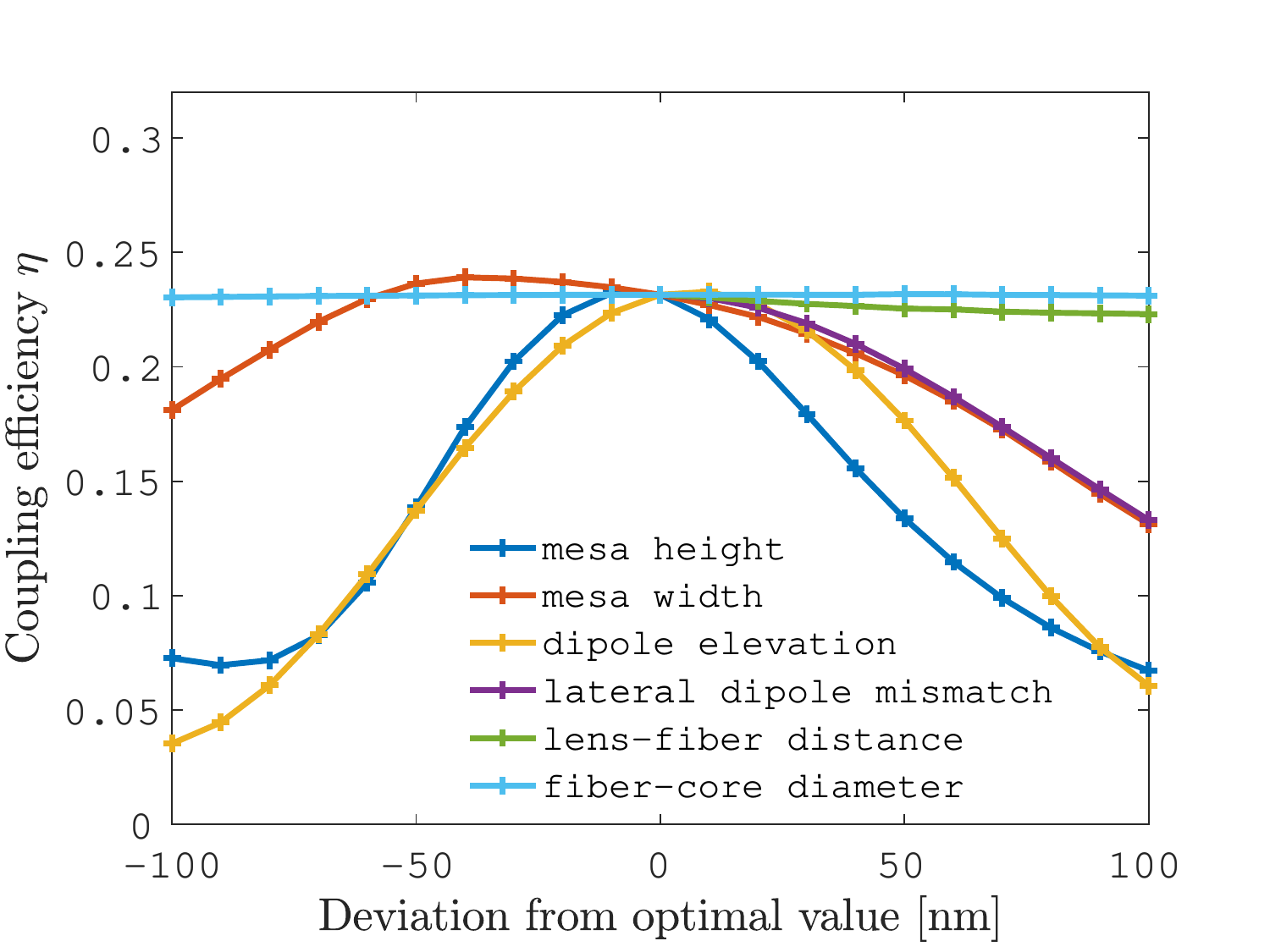} 
\vspace*{-4.8cm}
\\

\def\svgwidth{0.3\linewidth}
\phantom{.}\hspace{0.7cm}
\input{Sketch_spherical.pdf_tex}
\hspace{1.8cm}
\def\svgwidth{0.3\linewidth}
\input{Sketch_mesa.pdf_tex}

\vspace*{4.2cm}
\caption{Sensitivity of the coupling efficiency $\eta$ into a single-mode fiber with respect to deviations of the geometry parameters from their optimized values (see table~\ref{tab:optimal_parameters}) for the 
spherical lens setup (left) and the mesa setup (right).}
\label{fig:sensitivity}
\end{figure}

The sensitivities with respect to errors in the system parameters are depicted in Fig.~\ref{fig:sensitivity}. The figure shows that the spherical-lens setup is rather robust against fabrication errors of the lens width, the lens-fiber distance, the fiber-core diameter, and the lateral mismatch of the QD. On the other hand, an error in the lens height or the dipole elevation leads to a severe degradation of the coupling efficiency. 

In comparison, the coupling efficiency of the optimized mesa structure is more strongly influenced by fabrication errors of the width of the mesa and the lateral position of the QD (see right-hand side of Fig.~\ref{fig:sensitivity}). Also in this case, errors in the lens-fiber distance and the fiber-core diameter hardly influence the coupling efficiency. Altogether, both structures are relatively robust against lateral fabrication errors and thus meet current processing accuracy tolerances.

Noteworthy, the right-hand side of Fig.~\ref{fig:sensitivity} reveals, that the simple two-step optimization scheme did not converge to a local maximum. After the second optimization step (variation of the fiber-core diameter and the lens-fiber distance), the optimized values of the first step are not at the local maximum any more. For example, a lower value of the mesa width leads to a small increase of the coupling efficiency. 
Global optimization schemes, such as Bayesian optimization, tackle these problems by varying all system parameters over a given domain~\cite{SchneiderSantiagoetal.2017} and can thus further enhance the achievable extraction efficiency. However, the application of such methods is beyond the scope of this study.

The fabricated structures do not only suffer from errors of the geometry parameters, but also from an etching-induced surface roughness. Especially, the fabrication of spherical lenses can lead to non-negligible defects of the lens surface. The surface roughness breaks the cylindrical symmetry such that the corresponding effects can only be assessed by performing full 3D simulations. Therefore, we consider a reduced geometry, consisting of a Bragg reflector with only 10 instead of 25 double-layers. Although this reduces the optimal coupling efficiency to 19\%, the relative magnitude of roughness effects should be comparable between the full setup and the reduced one.

\begin{figure}
\begin{center}
\includegraphics[width=0.45\linewidth]{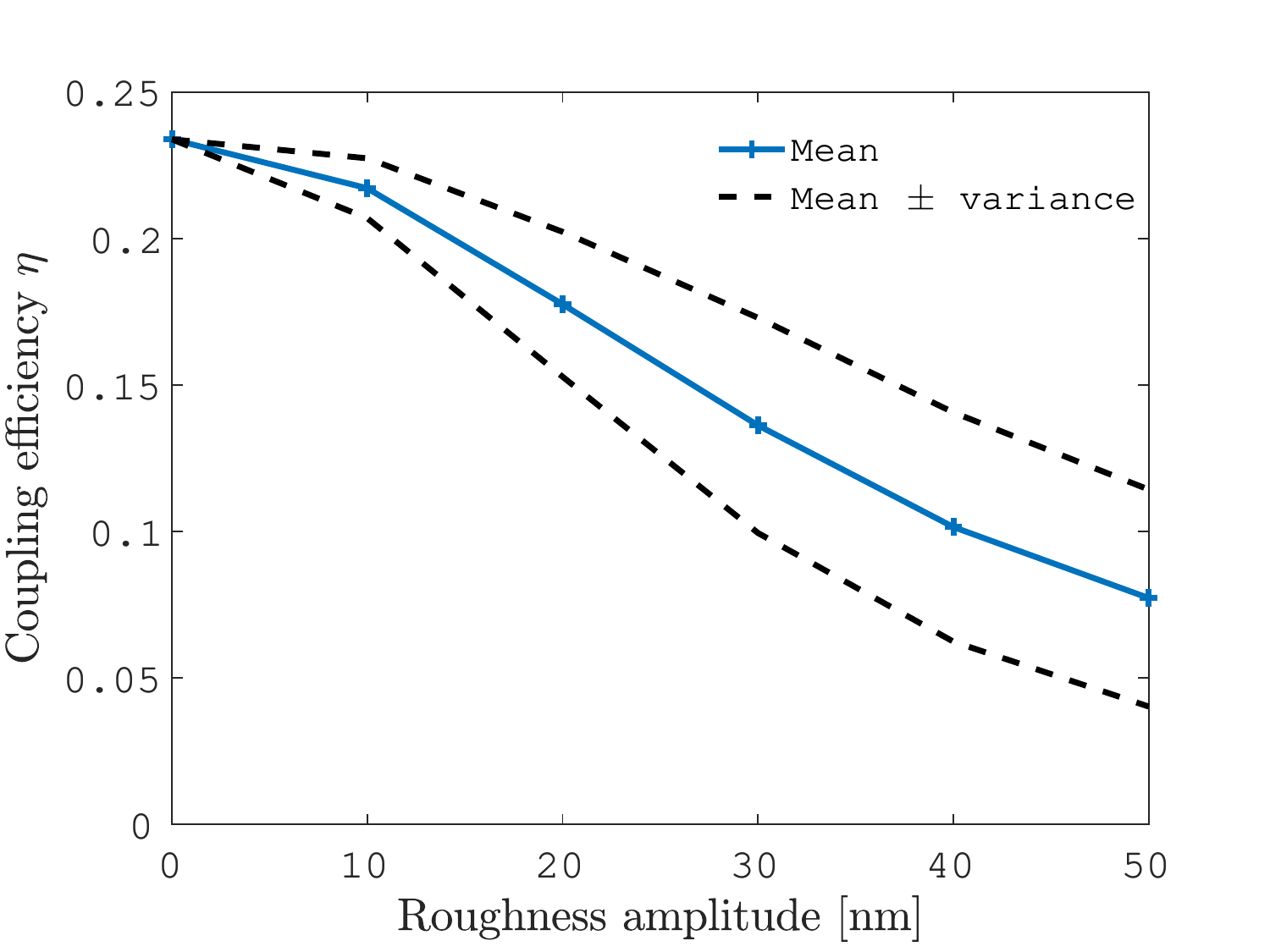} 
\def\svgwidth{0.45\linewidth}
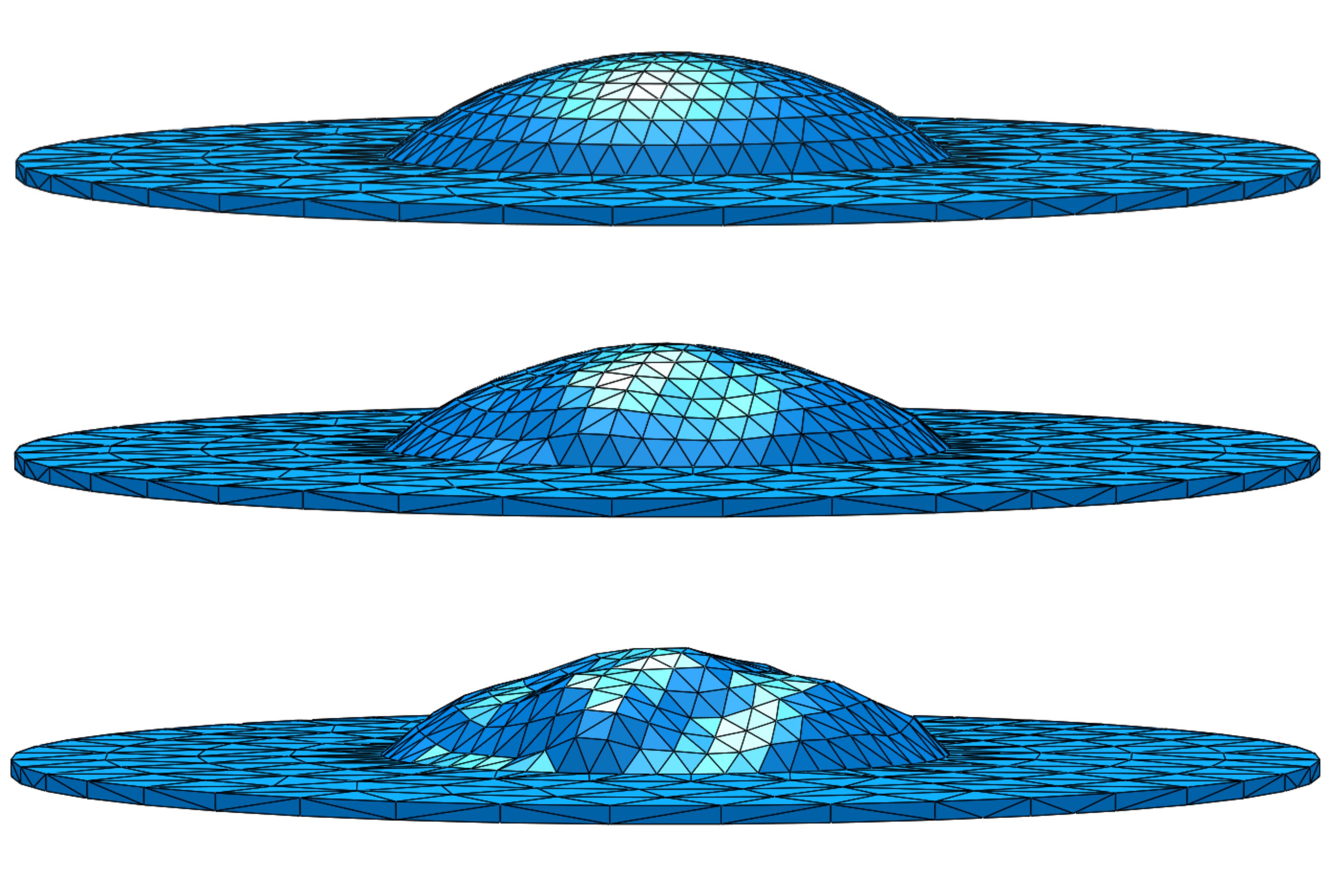
\end{center}
\caption{{\bf Left:} Mean and variance of coupling efficiency $\eta$ into a single-mode fiber for surface roughness of different amplitudes added to the surface of the optimized spherical lens. {\bf Right:} Visualization of the 3D mesh of the spherical lens for different roughness amplitudes.}
\label{fig:roughness}
\end{figure}

We assume that the surface roughness has a Gaussian autocorrelation. Hence, it can be parametrized by a single correlation length and a roughness amplitude~\cite{PhysRevLett.52.1798}. For the numerical study we vary the roughness amplitude between 0 and $50\,$nm and set the correlation length to $l=200\,$nm. The mean influence and the variance are estimated by generating 10 random roughness profiles for each amplitude. Figure~\ref{fig:roughness} shows how the coupling efficiency degrades with increasing roughness amplitude. With increasing amplitude also the variance of the degradation increases.
The numerical results show that surface roughnesses with an amplitude below 10~nm can be usually tolerated, while larger amplitudes quickly lead to a severe drop of the coupling efficiency.

\section{Conclusion}
\label{sec:Conclusion}

We have presented an efficient numerical method for the simulation of localized light sources embedded in micro-optical structures, taking into account both, the highly singular nature of the emitted light field as well as the rotational symmetry of the geometrical setup. The method was applied for studying the behavior of self-assembled quantum dots embedded into two types of diffractive structures -- spherical lens and mesa -- placed above a Bragg reflector.
We have performed a sampling of the parameter space of the system geometry and an optimization of the structure with respect to the coupling efficiency to the fundamental modes of a fiber. Moreover, we studied the robustness of the optimized geometries in the presence of fabrication errors.

The optimized geometries for the spherical lens and mesa setups presented here are not necessarily the global optima of the parameter space. Although the computation times are relatively small, a high-resolved grid search of the optimum in the full parameter space is still unfeasible. Global optimization schemes, such as Bayesian optimization, tackle this problem by sampling the parameter space efficiently based on a statistical model of the objective function~\cite{SchneiderSantiagoetal.2017}. For the future we plan to apply Bayesian optimization methods in order to extend the covered parameter space and to identify geometries with an even higher coupling efficiency.

\section{Appendix A -- dipole approximation}
In Coulomb gauge, the interaction between a charge distribution of a QD and the light field is determined by the Hamiltonian
\begin{equation}
\hat{\rm H}_{\rm int}(\mathbf{r}) = \hat{\mathbf{A}}(\mathbf{r}) \cdot \sum_n \frac{q_n}{m_n} i\hbar \nabla_{\mathbf{r}_n}, 
\end{equation}
where $\mathbf{A}$ is the vector potential of the electromagnetic field and $q_n$, $m_n$ and $\mathbf{r}_n$ are the charge, mass and position of the charges, respectively.

The dynamics of the system is governed by the coupling matrix elements of this operator in the basis of the Fock states of the electromagnetic field $\left|n\right>$ and the ground and excited state $\left|\Psi_g\right>$ and $\left|\Psi_e\right>$ of the QD. 
Within the dipole approximation one neglects the variation of $\mathbf{A}(\mathbf{r})$ over the extend of the charge distribution. That is, in the integration over the electronic degrees of freedom one replaces $\mathbf{A}(\mathbf{r})$ by $\mathbf{A}(\mathbf{r}_{\rm QD})$, where $\mathbf{r}_{\rm QD}$ is the position of the QD.
In this case the properties of the QD are described by its emission frequency $\omega$ and its dipole moment 
\begin{equation}
\mathbf{p} = - i\hbar \sum_n \frac{q_n}{m_n} \left<\Psi_g\left|\nabla_{\mathbf{r}_n}\right| \Psi_e \right>.
\end{equation}
Since the QD considered in this work has a larger horizontal than lateral extension the first state $\left| \Psi_e \right>$ is excited in a horizontal direction and the dipole moment in $z$-direction can be neglected.

The dipole approximation is only valid if gradients of the electric field can be neglected around the QD. Therefore, the wavelength in the embedding material $\lambda/n_{\rm struct}$ and the length scales of the surrounding geometry must be large compared to the dimensions of the QD. For a QD that is emitting with a vacuum wavelength of $1,300\,$nm we have $\lambda/n_{\rm struct}=382$~nm, which is an order of magnitude larger than the dimensions of the QDs (30~nm). Moreover, the comparison with experimental data shows that the dipole approximation provides reliable results as long as a QD is separated by more than roughly $100\,$nm from a material interface ~\cite{nphys1870,PhysRevLett.114.247401}.

If the QD is placed inside an optical cavity, the coupling strength between the QD and the light field can be described by the coupling constant \begin{equation}
\alpha = \frac{|\mathbf{p}|}{\hbar}\sqrt{\frac{\hbar\omega}{2\epsilon_0 V}},
\end{equation}
where $V$ is the volume of the cavity \cite{Novotny06principlesof}.
The strong coupling regime satisfies the condition $\alpha \gg \gamma$, where $\gamma$ is the photon decay rate inside the cavity. 
In this case the dynamical behavior of the quantum dot is strongly influenced by the surrounding cavity, such that only a QED treatment can accurately describe the system dynamics. A strong coupling leads, for example, to a frequency splitting of the emission spectrum. This effect cannot be reproduced by replacing the QD with a classical light source. 

For the optical structures considered in this study, the photon energy quickly radiates into the upper hemisphere. For this weak coupling regime ($\alpha \ll \gamma$) it can be shown that QED and the classical theory give the same results for the behavior of the spontaneous emission in the cavity\cite{Novotny06principlesof}. In this case the dipole can be described by a classical oscillating dipole moment $\mathbf{\mu}(t) = {\rm Re}\left\{\mathbf{p} e^{- i \omega t} e^{- \gamma t/2}\right\}$. 

The decay rate $\gamma$ inside a cavity is typically in the GHz regime. At optical frequencies the effects of the decay can be neglected, such that the QD can be modeled by an oscillating dipole with the point-like current density
\begin{equation}
\mathbf{J}(\mathbf{r},t) = \frac{d}{d t}\mathbf{\mu}(t) \delta(\mathbf{r}- \mathbf{r_{\rm QD}}) = \added{
- {\rm Re}\left\{ i \omega \mathbf{p} \delta(\mathbf{r}- \mathbf{r_{\rm QD}}) e^{- i \omega t}\right\}
}.
\end{equation}

\section{Appendix B -- waveguide mode orthonormality}

In the following we will show, that the eigenmodes of a waveguide \added{structure, such as an optical fiber,} are orthonormal with respect to the scalar product
\begin{equation}
\label{eq:scalar_prod2}
\left<\mathbf{E}_1,\mathbf{E}_2\right> = \frac{1}{2 i \omega}\int {\rm d}\mathbf{n} \cdot (\mathbf{E}_1 \times\mu^{-1}\nabla\times \mathbf{E}_2) 
= \frac{1}{2}\int {\rm d}\mathbf{n} \cdot 
  (\mathbf{E}_1 \times \mathbf{H}_2),
\end{equation}
where the integration is performed over the cross section of the waveguide (w.~l.~o.~g. at $z=0$).  

The eigenmodes $\mathbf{E}_n(\mathbf{r})$ for a given frequency $\omega$ are solutions of Eq.~\eqref{eq:time_harmonic} with no source current, i.~e.
\begin{equation}
\label{eq:mode_app}
\nabla\times\mu^{-1}\nabla\times \mathbf{E}_n(\mathbf{r}) - \epsilon\omega^2  \mathbf{E}_n(\mathbf{r}) = 0 .
\end{equation}

It can be easily confirmed that for any solution $\mathbf{E}_n$ also the field reflected at the $x-y$-plane at $z=0$
\begin{equation}
\widetilde{\mathbf{E}}_n(x,y,z) = \begin{pmatrix}
E_x(x,y,-z)\\ E_y(x,y,-z)\\ -E_z(x,y,-z)
\end{pmatrix}
\end{equation}
is a solution of of Eq.~\eqref{eq:mode_app}.

We consider the integral of the product of a reflected eigenmode $\widetilde{\mathbf{E}}_m$ and the left-hand side of Eq.~\eqref{eq:mode_app} over the volume along the waveguide from $z=0$ to $z=\delta$ 
\begin{equation}
\int_{\mathbb{R}^2 \times [0,\delta]} {\rm d V}\;\widetilde{\mathbf{E}}_m \cdot \left(\nabla\times\mu^{-1}\nabla\times \mathbf{E}_n - \epsilon\omega^2  \mathbf{E}_n\right) = 0
\end{equation}
We assume that all modes are guided and, hence, vanish for $r\rightarrow\infty$. In this case, two consecutive partial integrations of the volume integral yield
\begin{equation}
\label{eq:partial_integration}
\begin{split}
0 =& \int_{z=\delta} {\rm d A} \mathbf{e_z}\cdot( \widetilde{\mathbf{E}}_m \times\mu^{-1}\nabla\times \mathbf{E}_n) - \int_{z=0} {\rm d A} \mathbf{e_z} \cdot(\widetilde{\mathbf{E}}_m \times\mu^{-1}\nabla\times \mathbf{E}_n) \\
&-\int_{\mathbb{R}^2 \times [0,\delta]} {\rm d V}\; \mu^{-1}(\nabla\times\widetilde{\mathbf{E}}_m) \cdot (\nabla\times \mathbf{E}_n) - \epsilon\omega^2  \widetilde{\mathbf{E}}_m \cdot\mathbf{E}_n\\
=& \int_{z=\delta} {\rm d A} \mathbf{e_z}\cdot\left\{
  \widetilde{\mathbf{E}}_m \times\mu^{-1}\nabla\times \mathbf{E}_n +
  (\mu^{-1}\nabla\times \widetilde{\mathbf{E}}_m) \times \mathbf{E}_n
  \right\}\\
& - \int_{z=0} {\rm d A} \mathbf{e_z} \cdot \left\{
  \widetilde{\mathbf{E}}_m \times\mu^{-1}\nabla\times \mathbf{E}_n + 
  (\mu^{-1}\nabla\times \widetilde{\mathbf{E}}_m) \times \mathbf{E}_n
  \right\}\\
& - \int_{\mathbb{R}^2 \times [0,\delta]}  {\rm d V}\;\underbrace{(
  \nabla\times\mu^{-1}\nabla\times\widetilde{\mathbf{E}}_m
  ) \cdot \mathbf{E}_n 
  - \epsilon\omega^2  \widetilde{\mathbf{E}}_m \cdot\mathbf{E}_n.}_{=0}\\
\end{split}
\end{equation}
Since the waveguide is invariant along the $z$-direction, the eigenmodes have the general form of a plane wave
\begin{equation}
\mathbf{E}_n(x,y,z) = \mathbf{E}_n(x,y,0) e^{i k_n z}.
\end{equation}
Hence, the integration at $z=\delta$ differs from the one at $z=0$ only by a phase factor $e^{i(k_n-k_m)\delta}$. Furthermore,
partial integration of the surface integral yields
\begin{equation}
 \int {\rm d A} \mathbf{e_z} \cdot 
  (\widetilde{\mathbf{E}}_m \times\mu^{-1}\nabla\times \mathbf{E}_n)
=  
 \int {\rm d A} \mathbf{e_z} \cdot  
  (\mu^{-1}\nabla\times \widetilde{\mathbf{E}}_m) \times \mathbf{E}_n.
\end{equation}
Therefore, Eq.~\eqref{eq:partial_integration} simplifies to 
\begin{equation}
0=(e^{i(k_n-k_m)\delta}-1) \int_{z=0} {\rm d A} \mathbf{e_z} \cdot 
 (\widetilde{\mathbf{E}}_m \times\mu^{-1}\nabla\times \mathbf{E}_n).
\end{equation}

For $m\neq n$ this equation can only be satisfied if $ \int {\rm d A} \mathbf{e_z} \cdot 
  (\widetilde{\mathbf{E}}_m \times\mu^{-1}\nabla\times \mathbf{E}_n) = 0$. 
On the other hand, for $m=n$ we have the freedom to normalize the eigenmodes such that $\frac{1}{2i\omega}\int{\rm d A} \mathbf{e_z} \cdot (\widetilde{\mathbf{E}}_n \times\mu^{-1}\nabla\times \mathbf{E}_n) = 1$ (in units of Watts per square meter). 
Hence, it holds $\left<\widetilde{\mathbf{E}}_m,\mathbf{E}_n\right> = \delta_{n m}$.

Finally, we can omit the $z$-reflection (i.~e. replace $\widetilde{\mathbf{E}}_m$ by $\mathbf{E}_m$)
since the projection on $z$ depends only on the transverse field components (i.~e. $\mathbf{e}_z\cdot(\mathbf{a}\times \mathbf{b}) = a_x b_y - a_y b_x = \mathbf{e}_z\cdot(\mathbf{a}_\perp \times \mathbf{b}_\perp)$), which do not change under reflection along the $z$-axis. Altogether, this proves the orthonormality relation $\left<\mathbf{E}_m,\mathbf{E}_n\right> = \delta_{n m}$.

We note that for loss-free, non-active materials (${\rm Im}(\epsilon) =0$) the value of the scalar product is, up to a phase shift, the same as when using the usual Poynting vector based scalar product $\frac{1}{2}\int {\rm d}\mathbf{n} \cdot (\mathbf{E}_1 \times \mathbf{H}_2^*)$. However, the mode-orthonormality for damped materials holds only for the holomorphic form of Eq.~\eqref{eq:scalar_prod2}.

\section*{Funding}
Programm zur F\"orderung von Forschung, Innovationen und Technologien (Pro FIT) co-financed by the European Fund for Regional Development (EFRE) (FI-SEQUR 10160385); Deutsche Forschungsgemeinschaft (DFG) within the Sonderforschungsbereich 787 (SFB787).


\end{document}